\documentclass[aps,notitlepage,onecolumn,preprintnumbers,superscriptaddress,amsmath,amssymb,showpacs,nofootinbib,prd]{revtex4-1}

\usepackage{graphicx}
\usepackage{dcolumn}
\usepackage{bm}
\usepackage{multirow}

\usepackage{amssymb}
\usepackage{amsmath}
\usepackage{amsfonts}
\usepackage{url}
\usepackage{color}
\usepackage{xcolor}
\usepackage{xspace}
\usepackage{setspace}
\usepackage{slashed}
\usepackage{mathrsfs}

\renewcommand{\eqref}[1]{Eq.~(\ref{#1})}

\begin{document}
\preprint{UCI-TR-2017-13}

\title{Phenomenology of neutron-antineutron conversion}

\author{Susan Gardner}
\email{\texttt{gardner@pa.uky.edu}}
\affiliation{Department of Physics and Astronomy, University of Kentucky,
Lexington, Kentucky 40506-0055 USA}
\affiliation{Department of Physics and Astronomy, University of  California, Irvine, 
California 92697, USA}

\author{Xinshuai Yan}
\email{\texttt{xinshuai.yan@uky.edu}}
\affiliation{Department of Physics and Astronomy, University of Kentucky,
Lexington, Kentucky 40506-0055 USA}


\begin{abstract} 
We consider the possibility of neutron-antineutron ($n-{\bar n}$) 
conversion, in which the change of a neutron into an antineutron is mediated
by an external source, as can occur in a scattering process. 
We develop the connections between $n-{\bar n}$ conversion and 
$n-{\bar n}$ oscillation, in which a neutron spontaneously tranforms 
into an antineutron, noting that if  
$n-{\bar n}$ oscillation occurs in a theory with B-L 
violation, then 
 $n-{\bar n}$ conversion can occur also. 
We show how an experimental limit on $n-{\bar n}$ conversion 
could connect concretely to 
a limit on $n-{\bar n}$ oscillation, and vice versa, using effective field theory techniques
and baryon matrix elements computed in the M.I.T. bag model.
\end{abstract} 




\maketitle


\section{Introduction}

Establishing that the symmetry of baryon number minus lepton number, 
B-L, is broken in nature would demonstrate that dynamics beyond the 
standard model (SM) exists. This prospect is often 
discussed in the context of the origin of the neutrino mass, with B-L violation 
necessary both to make neutrinoless double $\beta$ (0$\nu \beta\beta$) decay possible
and to give the neutrino a Majorana mass~\cite{weinberg1979baryon,schechter1982neutrino}. 
The would-be mechanism of 0$\nu \beta\beta$ decay is unknown, so that it need not be realized 
through the long-range exchange of a Majorana neutrino, nor need it even utilize 
neutrinos at all --- yet its observation would imply 
the neutrino has a Majorana mass~\cite{schechter1982neutrinoless}. 
In this paper we discuss the complementary possibility of  
 B-L violation in the quark sector~\cite{mohapatra1980local,mohapatra1980phenomenology}, 
and we develop new pathways 
to its discovery 
through the consideration of $n$-${\bar n}$ conversion. 

We draw a distinction between an $n$-${\bar n}$ 
oscillation~\cite{kuzmin1970zhetf,Glashow1979,mohapatra1980local}, 
in which the neutron would 
spontaneously transforms into an antineutron with the same energy and momentum, and an 
$n$-${\bar n}$ conversion, in which the neutron would transform into an antineutron 
as mediated by an external source. 
The ability to observe $n$-${\bar n}$ oscillations is famously fragile and 
can disappear in the presence of ordinary matter and magnetic 
fields~\cite{Glashow1979,mohapatra1980phenomenology,cowsik1981some}. Such 
environmental effects impact the neutron and antineutron differently, 
in that the energies of a neutron 
and antineutron are no longer the same, 
so that one particle cannot convert spontaneously into the other and satisfy 
energy-momentum conservation, a constraint that unbroken Lorentz symmetry demands. 
It is technically possible, but experimentally involved, to remove matter and magnetic
fields to the extent that sensitive experimental searches with free neutrons 
become possible. 
The most recent, and most sensitive, of such 
experimental searches was completed more than twenty years ago~\cite{baldo1994new}, 
yielding a $n\bar n$ lifetime limit of $\tau_{n \bar n} = 0.85 \times 10^8\,\hbox{s}$ 
at 90\% confidence level (CL). 
A next-generation experiment is also under 
development~\cite{phillips2016neutron,milstead2015new}. Independently, searches for
neutron-antineutron oscillations in nuclei have been conducted, with the most stringent 
lower limit on the bound neutron lifetime being 
$1.9 \times 10^{32}$ years at 90\% CL 
for neutrons in $^{16}$O~\cite{Abe:2011ky}. 
Employing a probabilistic computation of the nuclear suppression 
factor~\cite{Dover:1982wv,Friedman:2008es}, with 
realistic nuclear optical potentials~\cite{Friedman:2008es}, 
yields an equivalent free neutron lifetime of 
$2.7\times 10^8\,\hbox{s}$ at 90\% CL~\cite{Abe:2011ky}. 
A recent study of the bound neutron lifetime
in deuterium~\cite{Aharmim:2017jna} 
also employs a probabilistic 
framework~\cite{Kopeliovich:2011aa} to determine that the equivalent free neutron lifetime 
is no less than $1.23 \times 10^8$ s at 90\% CL. 
We note that the ability of the free neutron experiment to observe 
a non-zero effect at its claimed 
sensitivity has been recently called into 
question~\cite{Kerbikov:2017spv}, due to the use of a probabilistic, rather than a 
quantum kinetic, 
framework for its analysis. 
We thus find it of particular interest to explore pathways to B-L violation for which 
these limitations do not apply. 

In this paper, we consider how it may be possible to observe B-L violation with baryons 
without requiring that a neutron spontaneously oscillates into an antineutron. 
One alternate path, that of dinucleon 
decay in nuclei~\cite{Mohapatra:1982eu,kabir1983limits,basecq1983deltab,arnold2013simplified}, 
is known and is being actively 
studied~\cite{berger1991lifetime,bernabei2000search,litos2014search,gustafson2015search}, 
though its 
sensitivity is limited by the finite density of bound nuclei. 
Another possibility occurs if the neutron transforms into an antineutron while 
coupling to an external vector current, as possible in a scattering process. 
The latter is not sensitive to the presence of matter and magnetic fields, 
because the external current permits energy-momentum
conservation to be satisfied irrespective of such effects. As we shall see, 
the leading-dimension effective operators that realize this are of higher mass dimension than 
those that give rise to $n$-${\bar n}$ oscillations, or dinucleon decay; 
however, the difference in mass
dimension need not be 
compensated by a new physics mass scale --- and thus 
the amplitude for an $n$-${\bar n}$ conversion need not be much smaller than that for an 
$n$-${\bar n}$ oscillation. 
Moreover, 
since the neutron is a composite of quarks, and the quarks carry both electric and color charge, 
operators that mediate $n$-${\bar n}$ oscillations can be related to those that
generate $n$-${\bar n}$ conversion. To make our discussion concrete, we consider 
the example of electron-neutron scattering, so that $n$-${\bar n}$ conversion would be mediated
by the electromagnetic current, though free neutron targets are not practicable. 
Rather, an ``effective'' neutron target such 
as $^2$H or $^3$He  
would be needed, though the neutron absorption cross section on $^3$He is too large to make
that choice practicable. The reactions of interest would thus include 
$e + ^2\!{\rm H} \to e + {\bar n} + X(n,p)$, or, alternatively, either 
$n + e \to {\bar n} + e$ or $n + ^2\!{\rm H} \to {\bar n} + p + X(e)$, 
where 
$X(n,p)$, e.g., denotes an unspecified final state containing a neutron and a proton. 
Studies with heavier nuclei, generally 
$n + A \to {\bar n} + X(n,p)$, with  $A$ denoting a nucleus such as $^{58}$Ni, 
could also be possible.

The observation of an $n$-${\bar n}$ transition would speak to 
new physics at the TeV scale, 
and particular model realizations contain not only 
TeV-scale new physics but also give neutrinos suitably-sized 
Majorana masses~\cite{mohapatra1980local,chacko1999supersymmetric,babu2001observable}. 
However, proving that the neutrino has a Majorana mass in the absence
of the observation of 0$\nu \beta\beta$ decay
requires not only 
an observation of B-L violation, but also that of another baryon number violating
process~\cite{babu2015determining}. 
Improving the experimental limit on the non-appearance of 
$n$-${\bar n}$ oscillations can also severely constrain particular models of 
baryogenesis~\cite{babu2009neutrino,babu2012coupling,babu2013post}. 
It could also help shed light on the mechanism of 0$\nu \beta\beta$ decay in nuclei, 
which could arise from a short-distance mechanism mediated by TeV-scale, B-L violating 
 new physics or a long-range exchange of a Majorana neutrino, with new physics appearing, rather, 
at much higher energy scales, as the supposed mechanism of 0$\nu \beta\beta$ decay. 
The continued non-appearence of neutron-antineutron oscillations and thus of 
TeV-scale physics may ultimately 
speak in favor of a light Majorana neutrino mechanism for 0$\nu\beta\beta$ decay. 

In this paper we develop the possibility of $n-{\bar n}$ conversion in a concrete way. 
We begin, in Sec.~\ref{effn}, by constructing an low-energy, effective Lagrangian in 
neutron and antineutron degrees of freedom with B-L violation, in which the hadrons 
also interact with electromagnetic fields or sources. It has been common to analyze the
sensitivity to $n-{\bar n}$ oscillations within an effective 
Hamiltonian framework~\cite{mohapatra1980phenomenology}; 
we employ its spin-dependent version~\cite{gardner2015} to 
show how the spin dependence of $n-{\bar n}$ conversion leaves the transition probability 
unsuppressed 
in the presence of a magnetic field. To 
redress the possibility of suppression from matter effects, however, a different
experimental concept is needed unless the matter were removed --- 
we refer to Ref.~\cite{Kerbikov:2017spv} 
for a discussion of the implied experimental requirements.
We develop the possibility of $n-{\bar n}$ conversion through
scattering in Sec.~\ref{quark}. 
Here our particular interest is how it might be connected 
theoretically to the possibility of 
$n$-${\bar n}$
oscillation.  
We do this by working at an energy scale high enough to resolve the quarks 
in the hadrons; thus to realize $n-{\bar n}$ conversion we start with the quark-level
operators that mediate $n-{\bar n}$ oscillation and dress the quarks with photons to enable 
electromagnetic scattering. With this we find the quark-level operators that mediate 
$n-{\bar n}$ conversion. In Sec.~\ref{connect} we compute the matching conditions to the 
hadron-level effective theory, computing the needed matrix elements in the 
M.I.T. bag model~\cite{Chodos:1974pn,Chodos:1974je}. 
This gives us a concrete connection between $n-{\bar n}$ conversion and oscillation. 
Finally in Sec.~\ref{proposals} we analyze the efficacy of different experimental 
pathways to produce $n-{\bar n}$ conversion, and particularly the best indirect limit
on $n-{\bar n}$ oscillation parameters, and we conclude with a summary and outlook 
in Sec.~\ref{outlook}. In a separate paper we develop how best to discover 
$n$-${\bar n}$ conversion in its own right~\cite{svgxy_search17}. 

\section{Low-energy $n$-${\bar n}$ transitions with spin} 
\label{effn}

At low-energy scales we can regard neutrons and antineutrons as effectively elementary particles 
and realize $n-\bar{n}$ transitions 
through B-L violating effective operators in these degrees of freedom. 
Previously we have shown that the 
unimodular phases associated with the discrete symmetry transformations of Dirac fermions 
must be restricted in the presence of B-L violation; particularly, we have found that
the phase associated with CPT must be imaginary~\cite{gardner2016c}. We 
refer the reader to Appendix A for a summary of our definitions and conventions. 
The notion that Majorana particles, being their own antiparticles, 
have special transformation properties under 
CPT, CP, and C is long known, as are their implications for the interpretation
of  0$\nu \beta\beta$
experiments~\cite{Kayser:1983wm,Kayser:1984ge}. 
More generally, the existence of phase constraints associated with the discrete symmetry
transformations of
Majorana fields had already been noted by 
Feinberg and Weinberg~\cite{Feinberg:1959}, as well as by Carruthers~\cite{Carruthers:1971}, 
with these authors determining 
the phase restrictions associated with the C, CP, T, and TP transformations. 
Haxton and Stephenson~\cite{Haxton:1985am} have also analyzed 
the phase constraint associated with C for a pseudo-Dirac neutrino~\cite{Wolfenstein:1981kw}, 
the case
most similar to that of the neutron, though they did not analyze 
the phase constraints associated with the other discrete symmetries. 
Under our CPT phase constraint, there are two leading mass dimension, 
CPT even and  Lorentz invariant, $n-\bar{n}$ transition operators, 
namely, ${\cal O}_1= n^T C n + {\rm h.c.}$ and ${\cal O}_2= n^T C\gamma_5 n + {\rm h.c.}$  
A third operator appears if we admit an interaction with an external vector 
current $j^\mu$~\cite{berev15}: 
${\cal O}_3=n^T C\gamma^{\mu}\gamma_5 n j_{\mu} + {\rm h.c.}$ Note that a
 B-L-violating  interaction of form $in^T C\slashed{\partial} n + {\rm h.c.}$ vanishes
under the use of the equation of motion for a free Dirac field. 
With this in hand, we find that 
the effective Lagrangian of $n-\bar{n}$ conversion, mediated through electromagnetic interactions, 
is 
\begin{eqnarray}
{\mathcal{L}}_{\rm eff} = i{\bar n} {\slashed{\partial}} n - M {\bar n}n - 
\frac{1}{2} \mu_n {\bar n} \sigma^{\mu \nu} n F_{\mu \nu}
-  \frac{\delta}{2} (n^T C n + {\rm h.c.} ) - 
\frac{\eta}{2} ( n^T C \gamma^\mu \gamma^5 n j_\mu + {\rm {h.c.}} ) \,, 
\label{Leff}
\end{eqnarray}
where $n$ denotes the neutron field with mass $M$ and magnetic moment $\mu_n$, $j^\mu$ is the 
current associated with a spin 1/2 particle with electromagnetic charge $Qe$, noting $F^{\mu \nu}$ is
the electromagnetic tensor, and 
$\delta$ and $\eta$ are real constants. 
Using Maxwell equations with Heaviside-Lorentz conventions, we can 
replace the current $j^\nu \equiv \bar \psi \gamma^\nu \psi$ with fields via 
$Qe j^\nu = \partial_\mu F^{\mu \nu}$ as convenient. 
We have neglected the possibility of a $n^T C \gamma_5 n + {\rm h.c.}$ term, although 
CPT and Lorentz symmetry permits it~\cite{gardner2016c}, because it does not contribute
to $n-{\bar n}$ oscillations~\cite{gardner2015,berev15,Fujikawa:2016sft}. 
The presence of the external current $j^\mu$ makes $n-{\bar n}$ transitions with a 
flip of spin possible. To illustrate the efficacy of this, we now turn to the
computation of the $n -{\bar n}$ transition probability in the presence of a 
non-uniform magnetic field.

To compute the transition probability in an effective Hamiltonian framework with 
spin degrees of freedom~\cite{gardner2015}, 
 we must work out the $4 \times 4$ mass matrix ${\cal M}$ associated 
with Eq.~(\ref{Leff}). Using the 
$i\in |n(\mathbf{p},+)\rangle$, $|{\bar n}(\mathbf{p},+)\rangle$, 
$|{n}(\mathbf{p},-)\rangle$, $|{\bar n}(\mathbf{p},-)\rangle$ basis with $\mathbf{p}=0$, we compute the
matrix elements of the Hamiltonian $H$ associated with $\mathcal{L}_{\rm eff}$ and define the elements of ${\cal M}$
so that ${\cal M}_{ij} = \langle i | H | j \rangle /2M$.
Evaluating ${\cal M}_{ij}$ explicitly, 
we note that matrix elements associated with the $\eta$-dependent term are spin dependent. 
Although magnetic fields do act to suppress $n-\bar{n}$ oscillations mediated by the $\delta$ term in $\mathcal{L}_{\rm eff}$, 
the behavior of the $\eta$-dependent term 
is different. 
Suppose a magnetic field $\mathbf{B}$ is present. We choose the spin quantization axis so that 
$\mathbf{S}$ is aligned with $\mathbf{B}$ and the $\hat{\mathbf{z}}$ axis. Defining $\omega_0 \equiv |\pmb{\mu}_n| |\mathbf{B}|$ and 
$\pmb{\omega} \equiv \eta \mathbf{j}$ with $Qe j^\nu = \partial_\mu F^{\mu \nu}$, we find 
\begin{eqnarray}
\mathcal{M}=\begin{pmatrix}
M+\omega_0 & \delta + \omega_z & 0 & \omega_x-i\omega_y \\
\delta + \omega_z & M-\omega_0 & \omega_x-i\omega_y & 0 \\
0 & \omega_x + i\omega_y & M-\omega_0 & \delta -\omega_z \\
\omega_x+i\omega_y & 0 & \delta -\omega_z & M+\omega_0
\end{pmatrix}\,,
\label{Mmatrix}
\end{eqnarray}
where we have assumed that $\mathbf{j}$ and $\mathbf{B}$ are roughly constant. 
If $\mathbf{B}$ is non-uniform and depends on the transverse coordinates
$x$ and $y$, then $\omega_x$ and $\omega_y$ can both be non-zero, whereas $\omega_z$ will vanish in the absence of an electric field. 
Introducing $\omega_{xy}=\sqrt{\omega_x^2+\omega_y^2}$,  
we solve for the eigenvalues and
eigenvectors of Eq.~(\ref{Mmatrix}) (with $\omega_z=0$) to determine the 
 probability that an neutron with spin $s=+$ transforms into an antineutron with $s=\pm$. 
We find 
\begin{eqnarray}
{\cal P}_{n+\rightarrow \bar{n}+}&=& \frac{\delta^2}{\delta^2 + \omega_0^2} \sin^2(t \sqrt{\delta^2 + \omega_0^2}) \cos^2 (t \omega_{xy})
\approx \frac{\delta^2}{\omega_0^2} \sin^2 t \omega_0 \,,
\label{refnnbar}\\
{\cal P}_{n+\rightarrow \bar{n}-}&=&\sin^2 (t\omega_{xy}) \left( \cos^2(t \sqrt{\delta^2 + \omega_0^2}) 
+ \frac{\omega_0^2}{\delta^2 + \omega_0^2} \sin^2(t \sqrt{\delta^2 + \omega_0^2}) \right) \approx \sin^2(t \omega_{xy}) \,,
\end{eqnarray}
where the approximate result reports the probability to leading order in B-L violation. 
We observe that $\mathcal{P}_{n+\rightarrow \bar{n}+}$ 
is quenched by the magnetic field, in that it becomes negligibly small 
unless $t\omega_0 \ll 1$, which is 
not surprising since the spin is not flipped. In contrast, $\mathcal{P}_{n+\rightarrow \bar{n}-}$ 
is not suppressed. Although we have illustrated the utility of the 
$j^\mu$ term 
in a particular case, our conclusion holds more generally. In particular, we can probe $n$-${\bar n}$ conversion 
through a scattering process, so that the neutron and antineutron 
do not have to have the same energy and momentum. 
Consequently, we are no longer bound to the context of an oscillation framework, and 
the quenching problems arising from the presence of either matter or magnetic fields 
are completely solved. 
For future reference we note that free $n-{\bar n}$ searches are
conducted in the so-called quasifree limit, so that 
Eq.~(\ref{refnnbar}) can be approximated by $\delta^2 t^2$. 
Thus a limit on the free $n {\bar n}$ lifetime $\tau_{n \bar n}$ 
corresponds to a limit on $\delta$ via $\delta = \tau_{n \bar n}^{-1}$, so that 
the limit from the ILL experiment can be expressed as 
$\delta \le 5 \times 10^{-32}\,\hbox{GeV}$ at 90\% CL~\cite{baldo1994new}. 

In what follows we develop how limits from low-energy scattering experiments that would search for $n$-${\bar n}$ transitions
can connect concretely to limits from $n$-${\bar n}$ oscillation searches. Before so doing, we note that a 
oscillation search after the  manner of existing experiments~\cite{baldo1994new} 
could also set a limit on $\eta$ directly by utilizing non-uniform or non-stationary electromagnetic fields
to generate a nonzero $\omega_{xy}$~\cite{svgxy_search17}.  

\section{$n$-${\bar n}$ transition operators at the quark level}
\label{quark}

Considering the $n$-${\bar n}$ transition operators of Eq.~(\ref{Leff}) 
from the viewpoint of simple dimensional analysis, we see that 
the mass dimension of $\delta$, $[\delta]$, has $[\delta]=1$, whereas $[\eta]=-2$ since $[j^\mu]=3$. 
Since  $[\eta/\delta]=-3$, one might think that 
$n-\bar{n}$ conversion would be suppressed by an additional factor of $\Lambda_{NP}^3$, 
where $\Lambda_{NP}$ is the cutoff mass scale of new physics.     
This is not necessarily true because of the presence of other energy scales. 
To illustrate this explicitly, we need to develop the form of the 
$n-\bar{n}$ 
conversion operators at the quark-level. We do this by considering energy scales at which the 
quark structure of the nucleon becomes explicit but are still well below the nominal scale of new physics, 
$\Lambda_{\rm QCD} \lesssim  E \ll \Lambda_{\rm NP}$. 
In this way we can realize quark-level $n$-${\bar n}$ conversion operators through 
electromagnetic interactions, by 
dressing the quarks of the quark-level $n-\bar{n}$ oscillation operators with photons, since 
the participating quarks also carry electric charge. 

The effective Lagrangian for $n-\bar{n}$ oscillations at the QCD scale involves operators with six quark fields, 
and which thus have an associated coefficient of mass dimension ${-5}$. 
Since these operators are key to our work, 
we briefly summarize their important ingredients. Based on our earlier discussion of the nucleon-level operators, we expect 
the quark-level ``building blocks'' to have the structure $q^{T\alpha}_{1\,\chi} C q^{\beta}_{2\,\chi}$, where the numerical 
and Greek indices are flavor and color labels, respectively. We work, too, in a chiral basis
with $\chi \in {\rm  L, R}$ and note that each quark block appears as a chiral pair, since operators of mixed chirality always vanish. 
The final $n$-${\bar n}$ operators should be compatible with the hadrons' flavor content
and also 
be invariant under color symmetry, SU(3)$_c$. There are 
three ways of forming an SU(3) singlet from a product of six fundamental representations in SU(3)$_c$. 
However, in the case of quarks of a single generation, 
only two color tensors can occur~\cite{Rao:1983sd}, namely, 
\begin{eqnarray}
(T_s)_{\alpha\beta\gamma\delta\rho\sigma}&=&\epsilon_{\rho\alpha\gamma}\epsilon_{\sigma\beta\delta}+
\epsilon_{\sigma\alpha\gamma}\epsilon_{\rho\beta\delta}+
\epsilon_{\rho\beta\gamma}\epsilon_{\sigma\alpha\delta}+\epsilon_{\sigma\beta\gamma}\epsilon_{\rho\alpha\delta}\,, \\
\label{Ts}
(T_a)_{\alpha\beta\gamma\delta\rho\sigma}&=&\epsilon_{\rho\alpha\beta}\epsilon_{\sigma\gamma\delta}+
\epsilon_{\sigma\alpha\beta}\epsilon_{\rho\gamma\delta}\, 
\label{Ta}
\end{eqnarray}
with $\epsilon$ denoting a totally antisymmetric tensor. 
We refer to Ref.~\cite{Rao:1983sd} for a discussion of B-L violating operators with arbitrary 
generational structure. 
Working in a chiral basis, so that $q_\chi \equiv (1 + {\chi} \gamma_5 ) q/2$ and
$\chi=\pm$ (or, equivalently, writing $q_\chi$ with $\chi={\stackrel{R}{{}_L}}$), 
we note, 
ultimately, that there are three types of $n$-${\bar n}$ operators~\cite{Rao:1982gt}:
\begin{eqnarray}
(\mathcal{O}_1)_{\chi_1\chi_2\chi_3}&=&[u^{T\alpha}_{\chi_1}C u^{\beta}_{\chi_1}]
[d^{T\gamma}_{\chi_2}C d^{\delta}_{\chi_2}][d^{T\rho}_{\chi_3}C d^{\sigma}_{\chi_3}](T_s)_{\alpha\beta\gamma\delta\rho\sigma}, 
\label{O1}
\\
(\mathcal{O}_2)_{\chi_1\chi_2\chi_3}&=&[u^{T\alpha}_{\chi_1}C d^{\beta}_{\chi_1}]
[u^{T \gamma}_{\chi_2}C d^{\delta}_{\chi_2}][d^{T\rho}_{\chi_3}C d^{\sigma}_{\chi_3}](T_s)_{\alpha\beta\gamma\delta\rho\sigma}, 
\label{O2} 
\\
(\mathcal{O}_3)_{\chi_1\chi_2\chi_3}&=&[u^{T\alpha}_{\chi_1}C d^{\beta}_{\chi_1}]
[u^{T \gamma}_{\chi_2}C d^{\delta}_{\chi_2}][d^{T\rho}_{\chi_3}C d^{\sigma}_{\chi_3}](T_a)_{\alpha\beta\gamma\delta\rho\sigma}\,,
\label{O3}
\end{eqnarray}
although only 14 of these 24 operators are independent, because the antisymmetric tensors yield the 
relationships~\cite{Rao:1982gt} 
\begin{eqnarray}
(\mathcal{O}_1)_{\chi_1LR}=(\mathcal{O}_1)_{\chi_1RL}\,, \ \ (\mathcal{O}_{2,3})_{LR\chi_3}=(\mathcal{O}_{2,3})_{RL\chi_3}\,, 
\label{nnbaroprel}
\end{eqnarray}
and~\cite{Caswell:1982qs}
\begin{eqnarray}
(\mathcal{O}_2)_{mmn}-(\mathcal{O}_1)_{mmn}=3(\mathcal{O}_3)_{mmn}\,,
\label{lastcon}
\end{eqnarray}
where $m, n \in [L,R]$. 
If we also demand that the operators be invariant under SU(2)$_L\times$U(1)$_Y$, the electroweak 
gauge symmetry of the SM, then finally only 
four operators are independent~\cite{Rao:1982gt,Caswell:1982qs}. For example, 
\begin{eqnarray}
&&\mathscr{P} _1=(\mathcal{O}_1)_{RRR}, \\
&&\mathscr{P} _2=(\mathcal{O}_2)_{RRR}, \\
&&\mathscr{P} _3=[q^{T i\alpha}_{L}C q^{j\beta}_{L}]
[u^{T\gamma}_{R}C d^{\delta}_{R}][d^{T\rho}_{R}C d^{\sigma}_{R}]\epsilon_{ij}(T_s)_{\alpha\beta\gamma\delta\rho\sigma}\nonumber \\
&&\ \ \ \ \ =2(\mathcal{O}_3)_{LRR}\,, \\
&&\mathscr{P} _4=[q^{T i\alpha}_{L}C q^{j\beta}_{L}]
[q^{T k\gamma}_{L}C q^{l\delta}_{L}][d^{T\rho}_{R}C d^{\sigma}_{R}]\epsilon_{ij}\epsilon_{kl}(T_a)_{\alpha\beta\gamma\delta\rho\sigma}\nonumber \\
&&\ \ \ \ \ =4(\mathcal{O}_3)_{LLR}\,, 
\end{eqnarray} 
where the Roman indices label the members of a left-handed SU(2) doublet. 

The matrix elements of these operators have been evaluated in the M.I.T. bag model by 
Rao and Shrock~\cite{Rao:1982gt} and, much more recently,  
in lattice QCD~\cite{Buchoff:2012bm, Syritsyn:2016ijx}. 
Once we have developed the quark-level $n$-${\bar n}$ conversion operators we, too, 
use the M.I.T. bag model to evaluate their matrix elements. We discuss noteworthy 
technical aspects of this in  Appendix B.

\subsection{From quark-level operators for $n$-${\bar n}$ oscillation to $n$-${\bar n}$ conversion}

\begin{figure}
\centering
\includegraphics[scale=0.55]{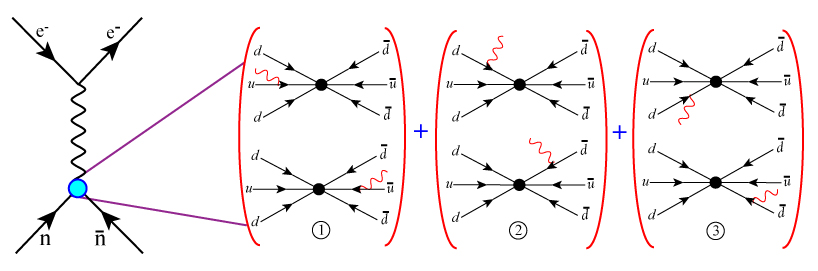}
\caption{A neutron-antineutron transition is realized through electron-neutron scattering. The virtual photon emitted from the 
scattered electron interacts with a general six-fermion $n-\bar{n}$ oscillation vertex. The particular graphs showns illustrate the 
two possible ways of attaching a photon to each of the blocks that appear in the $({\cal O}_1)_{\chi_1\chi_2\chi_3}$ operator of
Eq.~(\ref{O1}).}
\label{fig:QCD}
\end{figure}

Since dimensional analysis shows that the effective operator for $n$-${\bar n}$ conversion 
would be suppressed with respect to that for $n$-${\bar n}$ oscillation by three powers of a new-physics
mass scale, we wish to explore the manner in which we can use SM physics to 
find a more favorable relationship. In particular, since the quarks carry electric charge, we explore the possibility that the
external source in the $n$-${\bar n}$ conversion operator is the electromagnetic current. 
Of course quarks also carry color charge, but the associated current 
$\partial^{\mu} F_{\mu \nu}^a$ is not SU(3)$_c$ gauge invariant. 
In what follows we consider each of the $n$-${\bar n}$ transition operators in turn and determine the 
low-energy effective operator that follows from evaluating how its quarks interact with a virtual photon
generated by a scattered charged particle, such as an electron. In any particular, leading-dimension  $n$-${\bar n}$ operator, 
there are three blocks,  
and in each block there are 
two charged particles.
When a virtual photon is attached to these blocks, 
there are six possible ways that correspond to six different Feynman diagrams, as shown in Fig.~\ref{fig:QCD}. 
Note that we do not attach a photon line to the solid ``blob'' at the center because, as we shall see, 
this would yield an effect that would be suppressed by higher powers of the new physics mass scale. 

To determine the operator structures that emerge upon including electromagnetic 
interactions, we first compute the matrix element for
 the process $q^\rho (p) + \gamma (k) \to \bar{q}^\delta (p^\prime)$, noting that the
superscripts are flavor indices. 
Working in a chiral basis, the pertinent terms in the 
interaction Hamiltonian are 
\begin{equation}
{\cal H}_I \supset \frac{\delta_q}{2} \sum_{\chi_1} (\psi^{\rho\, T}_{\chi_1} C \psi^\delta _{\chi_1}  
+ {\bar \psi}^{\delta}_{\chi_1} C {\bar\psi}^{\rho\, T}_{\chi_1} )
+ Q_{\rho} e \sum_{\chi_2} \bar \psi_{\chi_2}^\rho \slashed{A} \psi_{\chi_2}^\rho + 
 Q_{\delta} e \sum_{\chi_3} \bar \psi_{\chi_3}^\delta \slashed{A}\psi_{\chi_3}^\delta \,,
\label{sampleham}
\end{equation}
where both $q^\rho$ and ${\bar q}^\delta$ have mass $m$. Computing 
\begin{equation}
\langle {\bar q}^\delta (p^\prime) | \mathcal{T} \left( \sum_{\chi_1, \chi_2}
\left( -i\frac{\delta_q}{2} \int d^4 x \psi^{\rho\, T}_{\chi_1} C \psi^\delta _{\chi_1} \right)
\left(-iQ_\rho e  \int d^4 y \bar \psi_{\chi_2}^\rho \slashed{A} \psi_{\chi_2}^\rho  
-iQ_\delta e  \int d^4 y \bar \psi_{\chi_2}^\delta \slashed{A}\psi_{\chi_2}^\delta\right)
\right) | q^\rho (p) \gamma (k) \rangle \,, 
\end{equation}
using standard techniques~\cite{Peskin:1995ev}, noting 
$\mathcal{T}$ is the time-ordering operator and the quarks are treated as free fields, 
we find 
\begin{equation}
- \frac{\delta_q}{2} e m i \sum_{\chi_2} \left( 
 Q_\rho \frac{{\bar u}^\delta (\mathbf{p}^\prime ,s^\prime ) \slashed{\epsilon}(k) ( 1 +\chi_2\gamma_5) 
u^\rho (\mathbf{p}, s)}{{p^\prime}^2 - m^2}
- Q_\delta \frac{{\bar v}^\rho (\mathbf{p},s) \slashed{\epsilon}(k) ( 1 +\chi_2\gamma_5) 
v^\delta (\mathbf{p}^\prime, s^\prime)}{p^2 - m^2}
\right) (2\pi)^4 \delta^{(4)}(p^\prime - p - k) \,,
\label{feyncalc1}
\end{equation}
where $\epsilon_{\mu}$ is the polarization vector of the photon, or, finally, 
\begin{eqnarray}
- \frac{\delta_q}{2} e m i \sum_{\chi_2}&& \Bigg( 
{\bar u}^\delta (\mathbf{p}^\prime ,s^\prime ) \slashed{\epsilon}(k) 
u^\rho (\mathbf{p}, s )\left(\frac{Q_\rho}{{p^\prime}^2 - m^2} 
- \frac{Q_\delta}{{p}^2 - m^2} \right) \nonumber \\
&&+ \chi_2 
{\bar u}^\delta (\mathbf{p}^\prime ,s^\prime ) \slashed{\epsilon}(k) \gamma_5 
u^\rho (\mathbf{p}, s )\left(\frac{Q_\rho}{{p^\prime}^2 - m^2} 
+ \frac{Q_\delta}{{p}^2 - m^2} \right) \Bigg) (2\pi)^4 \delta^{(4)}(p^\prime - p - k) 
\label{feyncalc2}
\end{eqnarray}
where we have employed the conventions and relationships of Appendix A throughout. 
Since $p^2 = p^{\prime\, 2}$, we see the 
vector term vanishes if $Q_\rho=Q_\delta$, as we would expect from CPT 
considerations~\cite{gardner2016c}. However, if $Q_\delta \ne Q_\rho$ 
the final result is non-zero even after summing over $\chi_2$. 
Replacing $\epsilon_\mu (k)$ with $k_\mu$ we see that the Ward-Takahashi identity is satisfied
after summing over $\chi_2$. For fixed $\chi_2$ the identity also follows once we sum over the photon-quark 
contributions that would yield an electrically neutral initial or final state, as in the case 
of $n-{\bar n}$ transitions. 
Thus we extract 
the effective operator associated with the quark-antiquark-photon vertex as 
\begin{equation}
-\frac{ m \delta_q e}{ p^2 - m^2}   \left( Q_{\rho} \psi^{\delta\,T}_{-\chi_2} C \gamma^\mu \psi^{\rho}_{\chi_2}
- Q_{\delta} \psi^{\delta\,T}_{\chi_2} C \gamma^\mu \psi^{\rho}_{-\chi_2} \right) \,, 
\end{equation}
noting that only the $C\gamma^\mu \gamma_5$ Lorentz structure would 
survive if $\rho=\delta$. For use in the 
neutron case we recast this as 
\begin{equation}
-\frac{ m \delta_q e}{ p^2 - m^2}   \left( Q_{\rho} \psi^{\delta\,T}_{-\chi} C \gamma^\mu \gamma_5 \psi^{\rho}_{\chi}
+ Q_{\delta} \psi^{\delta\,T}_{\chi} C \gamma^\mu \gamma_5 \psi^{\rho}_{-\chi} \right) \,, 
\label{qqbarconv}
\end{equation}
so that a sum over $\chi$ would yield the $C  \gamma^\mu \gamma_5$ Lorentz structure that appeared in 
our neutron-level analysis. Since we plan to study the $\chi$ dependence of the $n$-${\bar n}$ conversion
operator matrix elements, we may make this replacement 
without loss of generality. Studying the $\chi$ dependence 
reveals the 
interplay of the $C \gamma^\mu$ and $C \gamma^\mu \gamma_5$ Lorentz structures 
at the quark level, just as studying the $\chi_1$, $\chi_2$, and $\chi_3$ dependence 
of the $n$-${\bar n}$ oscillation matrix elements 
shows the interplay of $C$ and $C\gamma_5$ Lorentz structures, although only $C$ appears
in the neutron-level analysis. 
Since the quark is on its mass shell, we also have $p^2=m^2$, so that the 
explicit factor of $1/(p^2 - m^2)$ is problematic. However, the process 
we have computed ought not occur because it does not conserve electric charge. Rather, it may occur within 
a composite operator for which there is no change in electric charge, so that the 
participating quarks appear as part of a hadron state. 
Indeed the phenomenon of confinement in QCD reveals that quarks are never free, 
so that $p^2 - m^2$ does not vanish in the realistic case. 
We shall revisit its precise evaluation once the complete $n-{\bar n}$ conversion operator is in place. 

We can now proceed with our explicit construction of the $n$-${\bar n}$ conversion operator associated with 
$({\cal O}_1)_{\chi_1\chi_2\chi_3}$ in Eq.~(\ref{O1}), for which the pertinent Feynman graphs appear in Fig.~\ref{fig:QCD}. Using 
the effective vertex in Eq.~(\ref{qqbarconv}) for the two-quark-field-photon block, we see that the enumerated 
sets of graphs,
\textcircled{1}, 
\textcircled{2}, and \textcircled{3}, correspond to the effective vertices 
\begin{eqnarray}
&&\frac{-2e\delta_q }{3}\frac{m}{p^2-m^2} 
[u^{\alpha\,T}_{-\chi}C\gamma^{\mu}\gamma_5 u^{\beta}_{\chi} 
+ u^{\alpha\,T}_{\chi}C\gamma^{\mu}\gamma_5 u^{\beta}_{-\chi}]
[d^{\gamma\,T}_{\chi_2}C d^{\delta}_{\chi_2}][d^{\rho\,T}_{\chi_3}C d^{\sigma}_{\chi_3}]
(T_s)_{\alpha\beta\gamma\delta\rho\sigma} \, ,
\label{O1I} \\
&&\frac{+e\delta_q}{3}\frac{m}{p^2-m^2} 
[u^{\alpha\,T}_{\chi_1}C u^{\beta}_{\chi_1}]
[d^{\gamma\,T}_{-\chi}C\gamma^{\mu}\gamma_5 d^{\delta}_{\chi} 
+ d^{\gamma\,T}_{\chi}C\gamma^{\mu}\gamma_5 d^{\delta}_{-\chi}]
[d^{\rho\,T}_{\chi_3}C d^{\sigma}_{\chi_3}](T_s)_{\alpha\beta\gamma\delta\rho\sigma} \, ,
\label{O1II} 
\end{eqnarray}
and 
\begin{equation}
\frac{+e\delta_q}{3}\frac{m}{p^2-m^2} 
[u^{\alpha\,T}_{\chi_1}C u^{\beta}_{\chi_1}]
[d^{\gamma\,T}_{\chi_2}C d^{\delta}_{\chi_2}]
[d^{\rho\,T}_{-\chi}C\gamma^{\mu}\gamma_5 d^{\sigma}_{\chi} 
+ d^{\rho\,T}_{\chi}C\gamma^{\mu}\gamma_5 d^{\sigma}_{-\chi}]
(T_s)_{\alpha\beta\gamma\delta\rho\sigma} \, ,
\label{O1III} 
\end{equation}
respectively. 
Combining these vertices 
gives the effective operator generated from $({\cal O}_1)_{\chi_1\chi_2\chi_3}$, namely, 
\begin{eqnarray}
(\tilde{\mathcal{O}}_{1})^{\chi\, \mu}_{\chi_1\chi_2\chi_3}&=& 
\Big[
-2[u^{\alpha\,T}_{-\chi}C\gamma^{\mu}\gamma_5 u^{\beta}_{\chi} 
+ u^{\alpha\,T}_{\chi}C\gamma^{\mu}\gamma_5 u^{\beta}_{-\chi}]
[d^{\gamma\,T}_{\chi_2}C d^{\delta}_{\chi_2}][d^{\rho\,T}_{\chi_3}C d^{\sigma}_{\chi_3}] 
\nonumber \\
&+&
[u^{\alpha\,T}_{\chi_1}C u^{\beta}_{\chi_1}]
[d^{\gamma\,T}_{-\chi}C\gamma^{\mu}\gamma_5 d^{\delta}_{\chi} 
+ d^{\gamma\,T}_{\chi}C\gamma^{\mu}\gamma_5 d^{\delta}_{-\chi}]
[d^{\rho\,T}_{\chi_3}C d^{\sigma}_{\chi_3}] 
\nonumber \\
&+&
[u^{\alpha\,T}_{\chi_1}C u^{\beta}_{\chi_1}]
[d^{\gamma\,T}_{\chi_2}C d^{\delta}_{\chi_2}]
[d^{\rho\,T}_{-\chi}C\gamma^{\mu}\gamma_5 d^{\sigma}_{\chi} 
+ d^{\rho\,T}_{\chi}C\gamma^{\mu}\gamma_5 d^{\sigma}_{-\chi}]
\Big]
(T_s)_{\alpha\beta\gamma\delta\rho\sigma} \,. 
\label{O1conv}
\end{eqnarray}   
Finally, including the current term $Q ej^\mu(q)/q^2$ that appears through electromagnetic scattering, we have 
the effective $n-\bar{n}$ conversion operator  
\begin{eqnarray}
(\tilde{\mathcal{O}}_{1})^\chi_{\chi_1\chi_2\chi_3}=(\delta_{1})_{\chi_1\chi_2\chi_3}
\frac{e m}{3(p_{\rm eff}^2 -m^2)} \frac{Qe j_{\mu}}{q^2}
(\tilde{\mathcal{O}}_{1})^{\chi\, \mu}_{\chi_1\chi_2\chi_3}\,,
\label{O1convf}
\end{eqnarray} 
where $(\delta_{1})_{\chi_1\chi_2\chi_3}$ is the explicit low-energy constant associated with the 
$({\cal O}_1)_{\chi_1\chi_2\chi_3}$ operator and we replace $p^2 \to p_{\rm eff}^2$ for clarity
in later use. We now turn to the $n$-${\bar n}$ conversion 
operators associated with the other $n$-${\bar n}$ operators, 
$({\mathcal O}_2)_{\chi_1\chi_2\chi_3}$ and $({\mathcal O}_3)_{\chi_1\chi_2\chi_3}$ in 
Eqs.~(\ref{O2}-\ref{O3}). 
Although the block structure of these operators 
is quite different from $({\cal O}_1)_{\chi_1\chi_2\chi_3}$, determining 
the effective operators is nevertheless 
straightforward. Employing Eq.~(\ref{qqbarconv}) for the structure of each two-quark-photon 
block we find for 
the effective operator generated from $({\cal O}_2)_{\chi_1\chi_2\chi_3}$:
\begin{eqnarray}
(\tilde{\mathcal{O}}_{2})^{\chi\, \mu}_{\chi_1\chi_2\chi_3}&=& 
\Big[
[u^{\alpha\,T}_{-\chi}C\gamma^{\mu}\gamma_5 d^{\beta}_{\chi} 
- 2 u^{\alpha\,T}_{\chi}C\gamma^{\mu}\gamma_5 d^{\beta}_{-\chi}]
[u^{\gamma\,T}_{\chi_2}C d^{\delta}_{\chi_2}][d^{\rho\,T}_{\chi_3}C d^{\sigma}_{\chi_3}] 
\nonumber \\
&+&
[u^{\alpha\,T}_{\chi_1}C d^{\beta}_{\chi_1}]
[u^{\gamma\,T}_{-\chi}C\gamma^{\mu}\gamma_5 d^{\delta}_{\chi} 
- 2 u^{\gamma\,T}_{\chi}C\gamma^{\mu}\gamma_5 d^{\delta}_{-\chi}]
[d^{\rho\,T}_{\chi_3}C d^{\sigma}_{\chi_3}] 
\nonumber \\
&+&
[u^{\alpha\,T}_{\chi_1}C d^{\beta}_{\chi_1}]
[u^{\gamma\,T}_{\chi_2}C d^{\delta}_{\chi_2}]
[d^{\rho\,T}_{-\chi}C\gamma^{\mu}\gamma_5 d^{\sigma}_{\chi} 
+ d^{\rho\,T}_{\chi}C\gamma^{\mu}\gamma_5 d^{\sigma}_{-\chi}]
\Big]
(T_s)_{\alpha\beta\gamma\delta\rho\sigma} \,. 
\label{O2conv}
\end{eqnarray}   
The effective $n-\bar{n}$ conversion operator in this case is then 
\begin{eqnarray}
(\tilde{\mathcal{O}}_{2})^\chi_{\chi_1\chi_2\chi_3}=(\delta_{2})_{\chi_1\chi_2\chi_3}
\frac{e m}{3(p_{\rm eff}^2 -m^2)} \frac{Q e j_{\mu}}{q^2}
(\tilde{\mathcal{O}}_{2})^{\chi\, \mu}_{\chi_1\chi_2\chi_3}\,. 
\label{O2convf}
\end{eqnarray} 
Since $({\mathcal O}_3)_{\chi_1\chi_2\chi_3}$ has the same block structure as 
$({\mathcal O}_2)_{\chi_1\chi_2\chi_3}$, we can obtain its effective operator
by replacing $(T_s)_{\alpha\beta\gamma\delta\rho\sigma}$ 
by $(T_a)_{\alpha\beta\gamma\delta\rho\sigma}$ in Eq.~(\ref{O2conv}) to yield 
$(\tilde{\mathcal{O}}_{3})^{\chi\, \mu}_{\chi_1\chi_2\chi_3}$ and finally 
$(\tilde{\mathcal{O}}_{3})^\chi_{\chi_1\chi_2\chi_3}$ in analogy to 
Eq.~(\ref{O2convf}). 
The quantity 
$p^2 - m^2$ is effectively the quark ``off-shellness'' due to binding effects. 
We assess this by evaluating $E^2 - m^2$, where $E$ is the energy of the ground-state quark,
as determined in the M.I.T. bag model. 
We have checked that the $n$-${\bar n}$ matrix elements of 
all these effective operators satisfy the Ward-Takahashi identity. 
Barring the possibility of vanishing $n$-${\bar n}$ hadronic
matrix elements, we expect to have two $n$-${\bar n}$ conversion operators for every 
non-redundant $n$-${\bar n}$ oscillation operator. 

\begin{figure}
\centering
\includegraphics[scale=0.5]{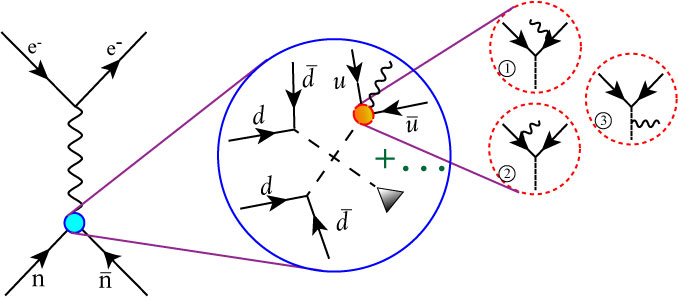}
\caption{A neutron-antineutron transition is realized through electron-neutron scattering in 
a particular model of B-L violation. 
We consider the case in which the virtual photon 
interacts with the neutron through a six-fermion $|\Delta B|=2$ vertex generated through the spontaneous 
breaking  of $(B-L)$-symmetry in the model of Ref.~\cite{mohapatra1980local}. 
A sample Feynman diagram is shown inside of the big blue circle, where the dashed line denotes 
a massive colored scalar. We represent the effective 
$q-\gamma-\bar{q}$ vertex by a red and gold circular area, which itself is realized 
by coupling the photon to any of the charged particles that appear.  
We explicitly show the diagrams that appear within three dashed red circles.}
 \label{fig5.3}
\end{figure}

We detour briefly to consider a particular model of B-L breaking, in order to demonstrate
that our low-energy, effective-operator analysis does indeed characterize the physics at 
leading power in the new-physics scale.
We pick a popular model in which $n-\bar{n}$ oscillations are generated 
through spontaneous breaking of a local $B-L$ symmetry associated with the ``partial unification'' group 
$\text{SU(2)}_L\otimes\text{SU(2)}_R\otimes\text{SU(4}'\text{)}$~\cite{mohapatra1980local}, where 
SU(4$'$) breaks to SU(3)$_{\rm c}\times$ U(1)$_{\rm B-L}$
at lower energies. 
A sample Feynman diagram of a  $|\Delta B|=2$ vertex, 
after Fig.~1 of Ref.~\cite{mohapatra1980local}, along with the three diagrams associated with the 
$u-\gamma-{\bar u}$ vertex, are shown in Fig.~\ref{fig5.3}. 
The pertinent terms of the interaction Hamiltonian can now be written as 
\begin{equation}
{\cal H}_I \supset 
\lambda \sum_{\chi_1} (\psi^{\rho\, T}_{\chi_1} C \psi^\delta _{\chi_1}  
\Delta_{\chi_1} + {\rm h.c.} )
+ Q_{\rho} e \sum_{\chi_2} \bar \psi_{\chi_2}^\rho \slashed{A} \psi_{\chi_2}^\rho + 
 Q_{\delta} e \sum_{\chi_3} \bar \psi_{\chi_3}^\delta \slashed{A}\psi_{\chi_3}^\delta 
+ Q_{\Delta_{\chi_1}}e \sum_{\chi_1} ( \Delta_{\chi_1}  A_\mu \partial^\mu \Delta_{\chi_1} + {\rm h.c.} ) \,,
\end{equation}
where $\Delta_{\chi_1}$ is a real scalar of mass $M$. 
Computing the three 
diagrams for 
 $q^\rho(p) + \gamma(k) \to {\bar q}^\delta (p^\prime)$, including $\Delta(k')$  as an 
intermediate propagator, we find
\begin{eqnarray}
\!\!\!\!\! - \frac{\lambda}{4} e i \sum_{\chi_1,\chi_2} \Bigg( && Q_\rho {\bar u}^\delta (\mathbf{p}^\prime ,s^\prime ) 
\frac{2m + \slashed{k'}(1-\chi_1 \gamma_5)}{(-p + k^\prime)^2 - m^2}
\slashed{\epsilon}(k) ( 1 +\chi_2\gamma_5) 
u^\rho (\mathbf{p}, s) 
\nonumber \\ && - Q_\delta {\bar v}^\rho (\mathbf{p},s) 
\frac{2m + \slashed{k'}(1-\chi_1 \gamma_5)}{(-p + k^\prime)^2 - m^2}
\slashed{\epsilon}(k) ( 1 +\chi_2\gamma_5) 
v^\delta (\mathbf{p}^\prime, s^\prime) \nonumber \\
&& 
+ 2 Q_{\Delta_{\chi_1}}  {\bar u}^\delta (\mathbf{p}', s') (1 + \chi_1 \gamma_5) u^\rho (\mathbf{p},s)
\frac{\epsilon_\mu (k) (2p^\mu - 2 p^\prime +k^\mu)}{(p-p')^2 - M^2} 
\Bigg)  \frac{1}{k^{\prime\,2} - M^2} (2\pi)^4 \delta^{(4)}(p^\prime + k^\prime - p - k) \,. 
\end{eqnarray}
Null results from collider searches for colored, scalar particles imply that 
$M$ can be no less than ${\cal O}(500\,{\rm GeV})$~\cite{Hayreter:2017wra}. 
In the low-energy limit, we thus have ${k'}^2  \ll M^2$, $(p - p')^2 \ll M^2$, 
and indeed $k'\to 0$. 
We see that the term in which a photon is radiated from a scalar is completely  
negligible in the low-energy limit, and the terms in which $\slashed{k'}$ appear are also negligible. 
Finally we thus recover the result of Eq.~(\ref{qqbarconv}) 
and the form of $({\cal O}_1)_{\chi_1 \chi_2 \chi_2}^{\chi\, \mu}$ we 
have found previously, noting $\lambda/M^6 = \delta/2$. 
Therefore, the particular model we have considered simply serves as 
a mechanism to generate the needed B-L violating interaction. 
Under the assumption that $n$-${\bar n}$ conversion is
mediated by electromagnetism, 
the possible conversion operators are determined 
 as long as one starts with a complete set of six-fermion $n-\bar{n}$ oscillation operators, 
irrespective of the 
model from which they arise.

\section{Matching from the quark to hadron level}
\label{connect}

Working in the quark basis, 
the effective Lagrangian that mediates $n-\bar{n}$ transitions without external sources is 
\begin{eqnarray}
{\cal L}_{\rm n / q}
\supset {\sum_{i,\chi_1,\chi_2,\chi_3}}^{\!\!\!\!\!\!\prime} (\delta_i)_{\chi_1,\chi_2,\chi_3}
({\cal O}_{i})_{\chi_1,\chi_2,\chi_3} + {\rm h.c.} \, ,
\label{quarknnbar}
\end{eqnarray}    
where $i=1, 2$, or $3$, and the prime denotes sums
restricted to yield a non-redundant operator set, as per 
the discussion in Sec.~\ref{quark}. We have already employed this form in determining
the $n$-${\bar n}$ conversion operators.
By analogy, the effective Lagrangian that mediates $n$-${\bar n}$ conversion is 
\begin{eqnarray}
{\cal L}^{\rm conv}_{\rm n / q} \supset 
\sum_\chi {\sum_{i, \chi_1,\chi_2,\chi_3}}^{\!\!\!\!\!\!\prime} (\eta_{i})^{\chi}_{\chi_1,\chi_2 ,\chi_3} 
(\tilde{{\cal O}}_i)^{\chi}_{\chi_1,\chi_2,\chi_3} 
+ {\rm h.c.}\,,
\end{eqnarray}  
though the precise nature of the restrictions in the sums requires further 
consideration. 
If the appearance of $n$-${\bar n}$ conversion derives from that of 
$n$-${\bar n}$ oscillation via electromagnetic interactions, then the 
$i, \chi_1, \chi_2, \chi_3$ sums are restricted as in Eq.~(\ref{quarknnbar}). Moreover, 
the low-energy constants should not depend on $\chi$, and the surviving terms follow
from computing the difference of $\chi=+$ and $\chi=-$. This in turn implies that
only some of the possible $n$-${\bar n}$ oscillation operators contribute to 
$n$-${\bar n}$ conversion. 
It is precisely this prospect that 
makes experimental searches for $n$-${\bar n}$ oscillations 
and $n$-${\bar n}$ conversion 
genuinely complementary. However, once we have operators of form
$(\tilde{{\cal O}}_i)^{\chi}_{\chi_1,\chi_2,\chi_3}$, 
broader possibilities follow. If we are agnostic as to their origin, then the redundacies
are those that follow from the flavor and color structure of the conversion operators themselves. 
Note that in this case we should also replace $Q_u/Q_d = -2$ in Eqs.~(\ref{O1conv})
and (\ref{O2conv}) by $g_{\rm ratio}$, an unknown parameter. 
We find, e.g., that relations of the form of Eq.~(\ref{nnbaroprel}) exist: 
\begin{eqnarray}
(\tilde{\mathcal{O}}_1)^{\chi}_{\chi_1LR}=(\tilde{\mathcal{O}}_1)^{\chi}_{\chi_1RL}\,, \ \ 
(\tilde{\mathcal{O}}_{2,3})^{\chi}_{LR\chi_3}=(\tilde{\mathcal{O}}_{2,3})^{\chi}_{RL\chi_3}\,. 
\label{nnbarconvrel}
\end{eqnarray}
Moreover, in the case of $(\tilde{{\cal O}}_1)^{\chi}_{\chi_1,\chi_2,\chi_3}$, 
the operator is also symmetric
under $\chi \to - \chi$. Beginning with 48 possible operators, we find, finally, that there are 
$6$ independent operators of form 
$(\tilde{{\cal O}}_1)^{\chi}_{\chi_1,\chi_2,\chi_3}$, and 12 independent operators of form 
$(\tilde{{\cal O}}_i)^{\chi}_{\chi_1,\chi_2,\chi_3}$ for each $i=2,3$, though the
$g_{\rm ratio}$ dependent terms should be separated into new operators if possible. 

In what follows, we relate the low-energy constants that appear in this Lagrangian to 
those in the low-energy Lagrangian 
at the nucleon level, in which, due to the low energy scale, we regard the neutron and 
anti-neutron as point-like particles. 
In particular, noting Eq.(\ref{Leff}), we 
relate the low-energy constants of this effective 
Lagrangian to those that appear at the quark level by equating the 
matrix elements of the pertinent operators. We have 
\begin{eqnarray}
\langle \bar{n} (\mathbf{p}^\prime ,s^\prime)|
\int d^3 x 
{\cal L}_{\rm eff}^{({\rm conv})} |n (\mathbf{p},s) \rangle
 =\langle \bar{n}_{\rm q} (\mathbf{p}^\prime ,s^\prime)|
\int d^3 x 
{\cal L}_{\rm n/q}^{({\rm conv})} 
|n_{\rm q} (\mathbf{p},s) \rangle \,,
\end{eqnarray}
where the 
states with the ``q'' subscripts are realized at the quark level. Explicitly, then,  
\begin{equation}
\delta {\bar v}(\mathbf{p}^\prime, s^\prime) C  u(\mathbf{p}, s)
=\langle \bar{n}_{\rm q} (\mathbf{p}^\prime ,s^\prime)| 
\int d^3 x 
{\sum_{i, \chi_1,\chi_2,\chi_3}}^{\!\!\!\!\!\!\prime} (\delta_{i})_{\chi_1,\chi_2 ,\chi_3} ({\cal O}_i)_{\chi_1,\chi_2,\chi_3} 
|n_{\rm q} (\mathbf{p},s) \rangle \,
\label{matchdelta}
\end{equation}
and
\begin{equation}
\eta {\bar v}(\mathbf{p}^\prime, s^\prime) C {\slashed{j}} \gamma_5 u(\mathbf{p}, s)
=\langle \bar{n}_{\rm q} (\mathbf{p}^\prime ,s^\prime)| \int d^3 x 
\sum_\chi {\sum_{i, \chi_1,\chi_2,\chi_3}}^{\!\!\!\!\!\!\prime} (\eta_{i})^{\chi}_{\chi_1,\chi_2 ,\chi_3} (\tilde{{\cal O}}_i)^{\chi}_{\chi_1,\chi_2,\chi_3} 
|n_{\rm q} (\mathbf{p},s) \rangle \,. 
\label{matcheta}
\end{equation}
Using the connections we have derived in Eq.~(\ref{O1convf}) and in and after Eq.~(\ref{O2convf})
we can rewrite the latter as 
\begin{equation}
\eta {\bar v}({\mathbf p}^\prime, s^\prime) C {\slashed{j}} \gamma_5 u({\mathbf p}, s)
= \frac{em}{3(p_{\rm eff}^2 -m^2)} \frac{ Qe j_\mu}{q^2} 
\langle \bar{n}_{\rm q} (\mathbf{p}^\prime ,s^\prime)| 
\!\!\int \!d^3 x 
\!\!\!\!\!{\sum_{i, \chi_1,\chi_2,\chi_3}}^{\!\!\!\!\!\!\prime} \!(\delta_{i})_{\chi_1,\chi_2 ,\chi_3} 
[(\tilde{{\cal O}}_i)^{{\rm R}\,\mu}_{\chi_1,\chi_2,\chi_3} 
\!- (\tilde{{\cal O}}_i)^{{\rm L}\,\mu}_{\chi_1,\chi_2,\chi_3}]
|n_{\rm q} (\mathbf{p},s) \rangle \,, 
\label{matcheta2}
\end{equation}
so that setting limits on $\eta$ can also constrain the quark-level low-energy constants
associated with $n$-${\bar n}$ oscillations. We will determine that the operator
matrix elements associated with $i=1$ vanish, 
so that $n$-${\bar n}$ conversion can only 
probe some of the $n$-${\bar n}$ oscillation operators. 
In the matching relations 
we have assumed that the quark-level low-energy constants are evaluated at the matching scale, 
subsuming evolution effects from the weak to QCD scales. Note, too, that we assume that 
``$\delta_i$'' in Eqs.~(\ref{matchdelta}) 
and (\ref{matcheta2}) are the same irrespective of such effects. 
Considering the matching relation of Eq.~(\ref{matcheta2}), we see that for a fixed experimental 
sensitivity to $\eta$ the limit on  $(\delta_i)_{\chi_1\chi_2\chi_3}$ will be sharpest if
$q^2 \simeq 0$. Thus in evaluating the hadron matrix elements we wish to choose 
$\mathbf{p} \simeq \mathbf{p}^\prime$. 
In the next section we compute the 
pertinent quark-level $n$-${\bar n}$ matrix elements explicitly using the M.I.T. bag model, 
for which the most convenient choice of kinematics is 
$\mathbf{p} = \mathbf{p}^\prime =0$.

\section{Matrix-element computations in the M.I.T. bag model} 
  \begin{table}
 \caption{Dimensionless matrix elements $(I_i)^{\chi 3}_{\chi_1 \chi_2 \chi_3}$ 
of $n-\bar{n}$ conversion operators. 
The column ``EM'' denotes the matrix-element combination of $(\chi={\rm R})-
(\chi={\rm L})$. 
} 
 \centering 
 \begin{tabular}{c c c c c c c c c c c c}
 \hline\hline 
 &  $I_1$ & & & & $I_2$ & & & & & $I_3$ &   \\[0.5ex]
 $\chi_1 \chi_2 \chi_3$  & $\chi=R$ & $\chi=L$ &  EM & 
$\chi_1 \chi_2 \chi_3$  & $\chi=R$ & $\chi=L$  & EM   & 
$\chi_1 \chi_2 \chi_3$  & $\chi=R$ & $\chi=L$  & EM \\[0.3ex]
\hline 
\hline 
 RRR & 19.8 & 19.8 & 0  & RRR & -4.95 & -4.95 & 0 & RRR & 1.80 & -8.28 & 10.1 \\ 
 RRL & 17.3 & 17.3 & 0 & RRL & -2.00 & -9.02 & 7.02 & RRL & -1.07 & -8.81 &  7.74 \\
 RLR & 17.3 & 17.3 & 0  & RLR & -4.09 & -0.586 & -3.50 & RLR & 7.20 & 6.03 & 1.17 \\
 RLL & 6.02 & 6.02 & 0 & RLL & -0.586 & -4.09 & 3.50 & RLL & 6.03 & 7.20 & -1.17 \\ 
 LRR & 6.02 & 6.02 & 0 & LRR & -4.09 & -0.586 & -3.50 & LRR & 7.20 & 6.03 & 1.17 \\ 
 LRL & 17.3 & 17.3 & 0 & LRL & -0.586 & -4.09 & 3.50 & LRL & 6.03 & 7.20 &  -1.17\\ 
 LLR & 17.3 & 17.3 & 0  & LLR & -9.02 & -2.00 & -7.02  & LLR & -8.78 & -1.04 & -7.74 \\ 
 LLL & 19.8  & 19.8  & 0 & LLL & -4.95  & -4.95  & 0 & LLL & -8.28  & 1.80 & -10.1  \\ 
\hline 
\hline  
\end{tabular}
\label{table:dim}
\end{table}
Since the M.I.T. bag model is well known~\cite{Chodos:1974pn,Chodos:1974je}, 
we only briefly summarize the ingredients 
that are important to 
our calculation. In this model, the 
quarks and antiquarks are confined 
in a static, spherical cavity of radius $R$ by a bag pressure $B$, within which they 
obey the free-particle Dirac equation. 
We only need the ground-state solutions, which we denote as 
$u^s_{\alpha,0} (\mathbf{r})$ ($v^s_{\alpha,0} (\mathbf{r})$) for a quark (antiquark) of flavor $\alpha$. 
We present their form and comment on 
the proper definition of $v^s_{\alpha,0}$ in Appendix B.
The quantized quark field is given by 
\begin{eqnarray}
\psi^i_\alpha (\mathbf{r})= \sum_{n, s}[b^{i}_{\alpha s}(\mathbf{p}_n) 
u^s_{\alpha,n} (\mathbf{r}) + 
d^{i\dagger}_{\alpha s} (\mathbf{p}_n) 
v^s_{\alpha,n} (\mathbf{r})] \,,
\end{eqnarray}
where $i$ is a color index and $b^{i}_{\alpha s}$ ($d^{i\dagger}_{\alpha s}$) 
denotes a quark (antiquark) annihilation (creation) 
operator, for which the 
non-null anticommutation relations are 
\begin{eqnarray}
\{b^{i}_{\alpha s}(\mathbf{p}), b^{j\dagger}_{\beta s'}(\mathbf{p}')\}&=&\delta_{s s'}\delta_{ij}\delta_{\alpha\beta}\delta^{(3)}(\mathbf{p}-\mathbf{p}'), \\
\{d^{i}_{\alpha s}(\mathbf{p}), d^{j\dagger}_{\beta s'}(\mathbf{p}')\}&=&\delta_{s s'}\delta_{ij}\delta_{\alpha\beta}\delta^{(3)}(\mathbf{p}-\mathbf{p}') \,.
\end{eqnarray}
The normalized neutron and antineutron wave functions are given by 
\begin{eqnarray}
|n\uparrow \rangle 
&=&(1/\sqrt{18})\epsilon^{ijk}(b^{i\dagger}_{u\uparrow}b^{j\dagger}_{d\downarrow}-b^{i\dagger}_{u\downarrow}b^{j\dagger}_{d\uparrow})
b^{k\dagger}_{d\uparrow}|0\rangle , \\
|\bar{n}\uparrow 
\rangle &=&(1/\sqrt{18})\epsilon^{ijk}(d^{i\dagger}_{u\uparrow}d^{j\dagger}_{d\downarrow}-d^{i\dagger}_{u\downarrow}d^{j\dagger}_{d\uparrow})
d^{k\dagger}_{d\uparrow}|0\rangle \, 
\end{eqnarray}
where we use 
$|n \uparrow \rangle \equiv | n_{\rm q}(\mathbf{0}, +) \rangle$ for spin up (and similarly for
${\bar n}$) 
and, following Rao and Shrock~\cite{Rao:1982gt}, we write 
the particular matrix elements of interest to us as 
\begin{eqnarray}
\langle \tilde{{\cal O}}_{i}\rangle^{\chi\,\mu}_{\chi_1\chi_2\chi_3}&\equiv &\langle \bar{n}\uparrow|\int d^3 \mathbf{r} (\tilde{{\cal O}}_i)^{\chi\,\mu}_{\chi_1\chi_2\chi_3}|n\uparrow\rangle 
=\left[\frac{N_\alpha^6 }{(4\pi)^2 p_\alpha^{3}}\right](I_i)^{\chi\,\mu}_{\chi_1\chi_2\chi_3}\,.
\end{eqnarray}    
We note that $(I_i)^{\chi\,\mu}_{\chi_1\chi_2\chi_3}$ are dimensionless integrals, and we refer the reader to 
Appendix B for the form of the pre-factor and all other technical details. 
Picking the $z$ component for evaluation, we have 
 \begin{eqnarray}
 (I_1)^{\chi 3}_{\chi_1\chi_2\chi_3}&=&
-2\Big(B_{11}(\chi,\chi_2,\chi_3)+B_{11}(-\chi,\chi_2,\chi_3)\Big)
+\Big(B_{12}(\chi_1,\chi,\chi_3)+B_{12}(\chi_1,-\chi,\chi_3)\Big)\nonumber \\
 &+& \Big(B_{13}(\chi_1,\chi_2,\chi)+B_{13}(\chi_1,\chi_2,-\chi)\Big) \,;
\end{eqnarray}
\begin{eqnarray}
 (I_j)^{\chi 3}_{\chi_1\chi_2\chi_3}&=&
\Big(B_{j1}(\chi,\chi_2,\chi_3)-2B_{j1}(-\chi,\chi_2,\chi_3)\Big)
+\Big(B_{j2}(\chi_1,\chi,\chi_3)-2B_{j2}(\chi_1,-\chi,\chi_3)\Big)\nonumber \\
 &+& \Big(B_{j3}(\chi_1,\chi_2,\chi)+B_{j3}(\chi_1,\chi_2,-\chi)\Big)
 \end{eqnarray}
for $j=2,3$, where 
\begin{eqnarray}
B_{11}(\chi_1,\chi_2,\chi_3)&=&-8i\Big[I_a-\frac{4}{3}\chi_2\chi_3 I_b-8\chi_2\chi_3 I_c\Big], \\
B_{12}(\chi_1,\chi_2,\chi_3)&=&-8i\Big[-2 I_a-\frac{28}{3}\chi_1\chi_3 I_b-8\chi_1\chi_3 I_c\Big], \\
B_{13}(\chi_1,\chi_2,\chi_3)&=&-8i\Big[-2 I_a-\frac{28}{3}\chi_1\chi_2 I_b-8\chi_1\chi_2 I_c\Big],
\end{eqnarray}
\begin{eqnarray}
B_{21}(\chi_1,\chi_2,\chi_3)&=&-4i\Big[-\frac{1}{2} I_a+\frac{2}{3}\chi_2\chi_3 I_b+4(\chi_1\chi_3-\chi_1\chi_2) I_b+4\chi_2\chi_3 I_c\Big], \\
B_{22}(\chi_1,\chi_2,\chi_3)&=&-4i\Big[-\frac{1}{2} I_a+\frac{2}{3}\chi_1\chi_3 I_b+4(\chi_2\chi_3-\chi_1\chi_2) I_b+4\chi_1\chi_3 I_c\Big], \\
B_{23}(\chi_1,\chi_2,\chi_3)&=&-4i\Big[\frac{5}{2} I_a+\frac{26}{3}\chi_1\chi_2 I_b+4\chi_1\chi_2 I_c\Big],
\end{eqnarray}
and 
 \begin{eqnarray}
B_{31}(\chi_1,\chi_2,\chi_3)&=&-\frac{1}{3}B_{21}(\chi_1,\chi_2,\chi_3)
-4i\Big[-\chi_1\chi_2 I_d+4\chi_2\chi_3 I_e\Big], \\
B_{32}(\chi_1,\chi_2,\chi_3)&=&-\frac{1}{3}B_{22}(\chi_1,\chi_2,\chi_3)
-4i\Big[-\chi_1\chi_2 I_d+4\chi_1\chi_3 I_e\Big], \\
B_{33}(\chi_1,\chi_2,\chi_3)&=&-\frac{1}{3}B_{23}(\chi_1,\chi_2,\chi_3)
-4i(3\chi_1\chi_2 I_f),
 \end{eqnarray}
with 
\begin{eqnarray}
\!\!\!\!\!&I_a = \int ^{\xi_\alpha}_0 dx\, x^2 [j^2_0(x)-\tilde{j}_1^2(x)]^2\Big(j^2_0(x)-\frac{1}{3}\tilde{j}^2_1(x)
\Big) \,, \quad
&\!\!\!I_b = \int ^{\xi_\alpha}_0 dx\,  x^2 [j^2_0(x)-\tilde{j}_1^2(x)]j^2_0(x)\tilde{j}^2_1(x) \,, \\
\!\!\!\!\!\!\!\!&I_c = \int^{\xi_\alpha}_0 dx\,  x^2 j^2_0(x)\tilde{j}^4_1(x) \,, \quad 
&\!\!\!I_d = \int ^{\xi_\alpha}_0 dx\, x^2 [j^2_0(x)+\tilde{j}_1^2(x)][j^2_0(x)-\tilde{j}_1^2(x)]\Big(j^2_0(x)+
\frac{1}{3}\tilde{j}^2_1(x)\Big) \,,\\
\!\!\!\!\!&I_e = \int ^{\xi_\alpha}_0 dx\, x^2 [j^2_0(x)+\tilde{j}_1^2(x)]j^2_0(x)\tilde{j}^2_1(x) \,, \quad 
&\!\!\!I_f = \int ^{\xi_\alpha}_0 dx\, x^2 [j^2_0(x)+\tilde{j}_1^2(x)]^2\Big(j^2_0(x)-\frac{1}{3}\tilde{j}^2_1(x)\Big) \,,
\end{eqnarray}
noting $\tilde{j}_1(x)\equiv \epsilon_0 j_1(x)$. 
We have three input parameters, $m_u$, $m_d$, and $R$, which can be 
determined by fitting the hadron spectrum. For definiteness we use the same fits as employed by 
 Rao and Shrock~\cite{Rao:1982gt}, 
in which $m_u=m_d$. Picking the fit with nonzero quark mass, so that 
$m_u=m_d=0.108\, {\rm GeV}$ and $R= 5.59\, {\rm GeV}^{-1}$, we evaluate 
$N_\alpha^6/(4\pi)^2 p_\alpha^3=0.529\times 10^{-5}\,{\rm GeV}^6$ (noting $\xi_\alpha \simeq 2.281$), as well as 
$I_a=0.298$,  $I_b = 0.0344$, $I_c=0.0106$, $I_d=0.396$, $I_e = 0.0557$, and $I_f = 0.460$. 
Dropping 
the common factor of $i$, we 
evaluate 
  $(I_i)^{\chi 3}_{\chi_1\chi_2\chi_3}$ numerically  and report the results in Table~\ref{table:dim}. 
As a numerical check, we have also computed 
the matrix elements of the 
$n-{\bar n}$ oscillation operators $({\cal O}_i)_{\chi_1 \chi_2 \chi_3}$, 
and we reproduce Rao and Shrock's results 
up to an overall factor of $-i$~\cite{Rao:1982gt}. The pattern of these results, i.e., the
sign and relative sizes, agree with
recent lattice QCD results~\cite{Syritsyn:2016ijx}, though the latter are of larger absolute
size. 
We note that the results of Table~\ref{table:dim} bear out the symmetries of Eq.~(\ref{nnbarconvrel}) and 
the $\chi\to -\chi$ symmetry of $(\tilde{{\cal O}}_1)^{\chi}_{\chi_1,\chi_2,\chi_3}$. 
Consequently, upon making the matrix element combination $(\chi={\rm R}) - (\chi={\rm L})$, as 
would occur if 
the conversion operators appear only via electromagnetic interactions, the contributions from 
$(\tilde{{\cal O}}_1)^{\chi}_{\chi_1,\chi_2,\chi_3}$ vanish as expected. 
This underscores the complementarity of
$n$-${\bar n}$ conversion and oscillation searches. Note, moreover, if we were to 
restrict our consideration to the conversion operators that
stem from electromagnetism and 
SU(2)$_L\times$U(1)$_Y$ invariant $n$-${\bar n}$ oscillation operators, only the
terms from $({\cal O}_3)_{LRR}$ and $({\cal O}_3)_{LLR}$ would survive.

\section{Numerical Estimates and Experimental Prospects}
\label{proposals}

We now turn to the numerical evaluation of the matching relations 
in Eqs.~(\ref{matchdelta},\ref{matcheta},\ref{matcheta2}). In the 
kinematic limit for which $\mathbf{p} = \mathbf{p}^\prime =0$, 
the left-hand sides of Eq.~(\ref{matchdelta}) and Eqs.~(\ref{matcheta},\ref{matcheta2}) become 
$\delta \delta_{s,s'}$ and 
$\eta( j^z s \delta_{s,s'} + j^x \delta_{s, -s'} + i s j^y \delta_{s,-s'})$, 
respectively. 
As a numerical check, we have verified 
that our numerical matching procedure does not depend on the particular non-zero 
component we pick. 
Generally, a plurality of quark-level operators can contribute to either $n$-${\bar n}$ 
oscillation 
or conversion; however, if a single operator dominates each case, then the 
experimental limits from the two processes can potentially be compared directly. 
Choosing $\mu=z$ as in the last section and 
supposing for illustration that only 
$({\cal{O}}_3)_{LLR}$ operates through electromagnetism, 
we find 
\begin{eqnarray}
(\delta_3)_{LLR}\left((I_3)^{R 3}_{LLR} - (I_3)^{L 3}_{LLR}\right)
\frac{e}{3}\frac{m}{p_{\rm eff}^2-m^2}\frac{Q e j_z}{q^2}=\eta j_z \,,
\end{eqnarray}      
whereas $(\delta_3)_{LLR}\langle \mathcal{O}_3\rangle_{LLR}=\delta$. 
Combining these relations 
we can thus relate a would-be limit on $\eta$ to one on $\delta$, namely 
\begin{eqnarray}
\eta=\frac{\delta}{q^2}\frac{Q e^2}{3}\left( \frac{m}{p_{\rm eff}^2-m^2} \right)
\frac{\left((I_3)^{R 3}_{LLR} - (I_3)^{L 3}_{LLR} \right)}{\langle \mathcal{O}_3\rangle_{LLR}}\,,
\end{eqnarray}
so that with $\eta \equiv  \beta \delta Q e^2/q^2$ we have 
 \begin{eqnarray}
 \beta\equiv \frac{1}{3}\left( \frac{m}{p_{\rm eff}^2-m^2} \right) \frac{
\left((I_3)^{R 3}_{LLR} - (I_3)^{L 3}_{LLR} \right)}{\langle \mathcal{O}_3\rangle_{LLR}}=\frac{1}{3}\left( \frac{0.108}{(0.365)^2} \right) \frac{(-7.72)}{2.03}
\ \text{GeV}^{-1}\simeq -0.946 \ \text{GeV}^{-1} \,.
\label{beta}
 \end{eqnarray}
In evaluating the kinematic factors we replace $p_{\rm eff}^2 - m^2$ with 
$E_\alpha^2 - m^2 = p_\alpha^2$ 
as an estimate of a 
quark's off-shellness. Noting Table~\ref{table:dim}, we see that other operators 
can also be used as distinct choices, with the associated values of 
$\beta$ ranging to roughly ten times smaller or larger. Returning to Eq.~(\ref{matcheta2}) 
we see we can recast the connection to the $n$-${\bar n}$ oscillation 
parameters more generally by replacing $\delta$ with $\tilde \delta$, where 
\begin{equation}
\tilde{\delta} \equiv 
\left[\frac{\langle \mathcal{O}_3\rangle_{LLR}} {(I_3)^{R 3}_{LLR} - (I_3)^{L 3}_{LLR}}
\right]\!\!
{\sum_{\chi_1,\chi_2,\chi_3}}^{\!\!\!\!\!\!\prime} 
\!\!\left[
(\delta_{2})_{\chi_1,\chi_2 ,\chi_3} \!\left( (I_2)^{{\rm R}\,3}_{\chi_1,\chi_2,\chi_3} 
\!\!\!- (I_2)^{{\rm L}\,3}_{\chi_1,\chi_2,\chi_3} \right)
+ (\delta_{3})_{\chi_1,\chi_2 ,\chi_3} \!\left( (I_3)^{{\rm R}\,3}_{\chi_1,\chi_2,\chi_3} 
\!\!\!- (I_3)^{{\rm L}\,3}_{\chi_1,\chi_2,\chi_3} \right)
\right]
\end{equation} 
where we neglect any momentum dependence in the nucleon matrix element.

To study the possible limits on $\eta$ and hence $\tilde\delta$, we consider the process 
$n (p_n) + \ell (p_\ell) \rightarrow \bar{n} (p_n^\prime) + \ell (p_\ell^\prime)$,
where $\ell$ is a charged lepton, or, more generally, any electrically charged particle. 
We study the limits on $\eta$ that would arise if 
we were to neglect the connection to $n$-${\bar n}$ oscillations in a separate 
paper~\cite{svgxy_search17}. 
To make  numerical estimates of the sensitivity to $\tilde\delta$, 
we consider different possible combinations of beam and target --- and 
different energy regimes as well, though we restrict our considerations to center-of-mass energies
well below nucleon-antinucleon production threshold. 
In all cases, however, the best limits on $\tilde\delta$ emerge if we consider kinematics in which 
the squared momentum transfer $q^2$ is minimized. 

To evaluate the scattering process, we note that the 
lowest-order amplitude in our effective field theory is simply 
$i\mathcal{M}=\eta\bar{u}_\ell (p'_\ell)(\gamma^{\mu})u_\ell(p_\ell) 
v^{T}_n(p'_n)C\gamma_{\mu}\gamma^5u_n(p_n)$, so that the spin-averaged, absolute-squared
amplitude $\overline{|\mathcal{M}|^2}$ is 
\begin{equation} 
\overline{|\mathcal{M}|^2}
=\frac{8 Q_\ell^2 e^4|\beta|^2\tilde\delta^2}{q^4}\Big[(p'_\ell\cdot p'_n)(p_\ell\cdot p_n)
+(p'_n\cdot p_\ell)(p'_\ell\cdot p_n)
-2m_\ell^2 M^2+ M^2(p'_\ell\cdot p_\ell)-m_\ell^2(p'_n\cdot p_n)\Big],
\label{amplitudesquare}
\end{equation} 
where we introduce $q^2 \equiv (p_\ell - p'_\ell)^2$ and 
$\overline{|{\mathcal M}'|^2}$ for the quantity in square brackets
for
subsequent use. The differential cross section is 
\begin{eqnarray}
d\sigma =\frac{1}{F
}\frac{d^3p'_\ell}{(2\pi)^3}\frac{1}{2E'_\ell}\frac{d^3p'_n}{(2\pi)^3}\frac{1}{2E'_n}
\overline{|\mathcal{M}|^2}(2\pi)^4\delta^{4}(p_\ell+p_n-p'_\ell-p'_n),
\label{differential_cross}
\end{eqnarray}
where the flux factor $F$ is $4((p_n \cdot p_\ell)^2 - m_\ell^2 M^2)^{1/2}$. In what
follows we evaluate the cross sections for electron-neutron scattering in various kinematics. 
We employ beam parameters and target densities as established at existing experiments and 
facilties, or their planned extensions, to estimate the limits on $\tilde \delta$ 
that can emerge 
through $n$-${\bar n}$ conversion. 

We begin with the case of an electron beam scattering from a neutron bound in a 
deuterium target, e.g., because a free neutron target is unavailable. 
In such a scenario the converted ${\bar n}$ is left in situ, to annihilate with the 
material
around it. In the alternate case of a neutron beam scattering from an 
atomic target, the converted ${\bar n}$ emerges with roughly the same momentum as 
the incoming beam, so 
that the location of the annihilation products serves as a background discriminant. 
In nuclear stability studies, experimental backgrounds 
arise from the interactions of atmospheric neutrinos within the experimental volume, 
producing charged leptons and hadrons, 
and they 
have been studied in detail, noting Refs.~\cite{Abe:2011ky,Aharmim:2017jna}, e.g. 
Neutrinos could also 
potentially mediate the reaction ${\bar \nu}  {n} \to \bar{n} \nu$, which would 
seem to suggest that a $n \to {\bar n}$ process could appear 
without breaking B-L symmetry. 
However, this effect should be negligibly small because 
it contains the product of a $|\Delta B|=2$ and a $|\Delta L|=2$ transition. 
In contrast to nuclear stability studies, it has proved possible to conduct a 
sensitive experimental search for free $n$-${\bar n}$ oscillation such that 
no background events that would mimic the signal appear~\cite{baldo1994new}. 

\subsection{Electron scattering from a deuterium target}
\label{eonn}
We evaluate the cross section for electron-neutron scattering, neglecting
the effect of nuclear binding. Integrating over phase space using
Eqs.~(\ref{amplitudesquare}) and (\ref{differential_cross}), 
assuming the neutron is initially at rest, we have 
\begin{eqnarray}
\frac{d\sigma}{d \Omega} &=& \frac{e^4 |\beta|^2 {\tilde\delta}^2}{8\pi^2 M |\pmb{p}_e|} 
\left(\frac{|\pmb{p'}_e|^2}{|\pmb{p'}_e|(M+ E_e) -|\pmb{p}_e|E'_e\cos\theta}\right)
\frac{\overline{|{\mathcal M}'|^2}}{q^4}
\Big |_{p'_n=p_e + p_n - p'_e \,;\, E_e' = \chi_e } \,,
\label{nrestphase}
\end{eqnarray}
where $\theta$ is the angle between $\pmb{p'}_e$ and $\pmb{p}_e$ and 
$E_e' =\chi_e$ is the solution to 
\begin{equation}
2M(E_e - E_e') + 2m_e^2 - 2 E_e E_e' + 2 |\pmb{p}_e| \sqrt{E_e'^2 - m_e^2} \cos\theta =0  \,.
\label{chie}
\end{equation}
The cross section grows as $\theta^{-4}$ as $\theta$ approaches zero, so that we can assess its
size for fixed $\tilde \delta$ by first estimating 
the minimum value of $\theta_0$. Noting Ref.~\cite{Fernow:1986}, 
we determine $\theta_0$ in two different ways and then choose the larger value for our cross 
section estimate. (i) Using the Coulomb interaction between the incoming charged particle
and the charged particles of the deuterium target, we estimate
\begin{eqnarray}
\theta_0\approx  \frac{2 \alpha}{r_a |\pmb{p_e}|} = \frac{2 m_e \alpha^2}{|\pmb{p_e}|}
=5.45\times 10^{-7}\,\hbox{rad}\,,
\end{eqnarray}
where we use $|\pmb{p_e}|=100\,{\hbox{MeV}}$. 
(ii) Alternatively, 
by the uncertainty principle, hitting an atom of $r_a$ in transverse size implies 
the incident momentum is smeared by $1/r_a$, 
so that the associated minimum  scattering 
angle is 
\begin{eqnarray}
\theta_0\approx \frac{1}{|\pmb{p}_e| r_a}= \frac{m_e \alpha}{|\pmb{p}_e|}
=3.73\times 10^{-5}\,\hbox{rad}\,,
\label{thetau}
\end{eqnarray}
where the radius of the hydrogen atom is used for $r_a$. Thus we see that the two
angular estimates are not the same, and we proceed to estimate the total cross section as per
\begin{equation}
\sigma \approx \frac{d\sigma}{d\Omega}\Big |_{\theta=\theta_0} \times \pi \theta_0^2 \,,
\label{totalest}
\end{equation} 
and the larger $\theta_0$, 
so that our total cross section estimate is 
\begin{eqnarray}
\sigma \approx \left[ |\tilde \delta|^2 \ 5.12\times 10^7 \right] \ \text{GeV}^{-2} \,,
\end{eqnarray}
where $|\tilde\delta|$ is understood to be in units of GeV. 
We note that as long as $\pmb{p}_e \gg m_e$ and $\theta_0 \propto 1/|\pmb{p}_e|$, 
the estimate no longer depends on $|\pmb{p}_e|$. To determine 
the sensitivity to $\tilde\delta$, 
we must compute the event rate $dN/dt$ and finally the expected
yield of events. We have 
\begin{equation} 
\frac{d N}{d t} = {\cal L} \sigma = \phi \rho_T L \sigma \,,
\end{equation}
where the luminosity ${\cal L}$ is in units of particles/s-cm$^2$, 
$\phi$ is the flux in units of particles/s, $\rho_T$ is the target number density, and $L$
is its length. We turn to the DarkLight experiment operating at the Free-Electron Laser (FEL) 
facility of Jefferson National Accelerator Laboratory (JLab)  for suitable electron 
beam parameters~\cite{Corliss:2017tms,Alarcon:2013bca,Neil:2005jy}. 
The beam energy that experiment employs is $100$ MeV, and its current is 10 mA, for a beam 
power of 1 MW; it also uses a gaseous hydrogen target. 
In our case we would favor a liquid deuterium target, however, because it is denser, noting, 
e.g., its use in the experiment of Ref.~\cite{Ito:2003mr}, 
but target heating may preclude its use under DarkLight conditions. To consider this issue
further we turn to 
the Qweak experiment at JLab~\cite{Allison:2014tpu}, which uses a liquid hydrogen 
target and an electron beam with energy $E=1.16\ \text{GeV}$ and a beam current of 
$180\  \mu A$, yielding a beam power in this latter case of $0.209$  MW. 
Thus for the estimate in our case, we suppose that 
we can lower the beam energy to 20 MeV, but keep the same beam current, 
so that the beam power will be within the range of the Qweak experiment and 
a liquid deuterium target can be used. The electron beam flux in this case is 
$0.6 \times 10^{17} \,\text{s}^{-1}$, 
the number density of liquid deuterium is $\rho_d= 5.1 \times10^{22}\ \text{cm}^{-3}$ 
at 19 K~\cite{clusius}, and if we suppose a 1 m long target and a running time of 1 year, 
the sensitivity to $\tilde \delta$ for $N$ signal events is 
\begin{eqnarray}
|\tilde\delta| \lesssim 2 \times 10^{-15} \sqrt{\frac{N\ \text{events}}{1\ \text{event}}}
\sqrt{\frac{1\ \text{yr}}{\text{t}}}
\sqrt{\frac{0.6 \times 10^{17}\ \text{s}^{-1}}{\phi}}
\sqrt{\frac{1\ \text{m}}{L}}\sqrt{\frac{5.1 \times 10^{22}\ \text{cm}^{-3}}{\rho}}\ \text{GeV}.
\end{eqnarray}

\subsection{Neutron scattering from atomic targets}
\label{none}
We now turn to the evaluation of neutron scattering from the charged particles 
of an atomic target. 
In this case the scattering cross section is dominated by the 
electron contribution, and we can ignore the role of electromagnetic $n-p$ 
scattering completely. However, it is important to pick a target for which the
loss of neutrons from neutron capture would be minimal. In this regard, a deuterium 
target would be a good choice because the measured thermal neutron capture cross section 
on deuterium is merely $\sigma_{\rm c} = 0.508 \pm 0.015\ \text{mb}$~\cite{Jurney:1982zz}, so that 
 neutron capture effects would have a very limited impact on the transmitted flux. 
That is, noting that the neutron flux loss would be controlled by 
$\phi = \phi_0 \exp(-\sigma_c L \rho_d)$, we have $\phi/\phi_0 \simeq 0.998$ if $L=1\ \text{m}$. 
Although the capture cross section scales inversely with the neutron velocity at low energies, 
capture effects are also negligible for cold neutron energies, noting a kinetic energy of 
$T_n \approx 10^{-3}\ \text{eV}$ for reference. 
Thus we regard a deuterium target as a suitable choice, though there are others, notably one of 
$^{16}$O. Potentially one could also 
consider neutron scattering from an electron plasma, confined by a magnetic field, 
but in this case the electron density is limited~\cite{brillouin} --- and
much greater electon 
number densities can easily be found in atomic targets. Thus 
we focus on the use of deuterium and oxygen targets. 

To determine the differential cross 
section for neutron-electon scattering, we need only switch the roles of the 
neutron and electron in the evaluation of the phase space integrals in 
Eq.~(\ref{nrestphase}). Certainly the 
differential cross section is still largest in the forward direction, 
and we continue to use Eq.~(\ref{totalest}) to estimate the total cross section. 
Solving the analog of Eq.~(\ref{chie}) for $E_n'$ reveals 
that the neutron scattering angle is restricted, that is, 
$m_e^2 - M^2 \sin^2 \theta \geq 0$ is a necessary consequence of the kinematics. 
We use the largest allowed scattering angle in 
our $n-e$ cross section estimate, 
yielding $\theta_{\rm max} \leq m_e/M\approx 5.44 \times 10^{-4}\,\text{rad}$. 
In the case of $n-p$ scattering, the scattering angle is no longer restricted, and 
the most reasonable choice of angle depends on the 
momentum of the neutron beam. For neutrons with a kinetic energy of 
100 MeV,  
e.g., we find 
$|\pmb{p}_n| =\sqrt{2M T_n} \approx 447\,\text{MeV}$, and in this case 
we adapt our uncertainty principle estimate of Eq.~(\ref{thetau}) 
to write $ \theta_{\rm min} \approx 1/|\pmb{p}_n| r_a $ and use that angle in our estimate. 
For cold neutrons, noting the conditions of the experiment of Ref.~\cite{baldo1994new}, 
we consider a kinetic energy of $T_n = 2 \times 10^{-3} \ \text{eV}$, which 
corresponds to an average neutron wavelength $\lambda \approx 6.5$ {\AA} and 
$|\pmb{p}_n| \approx  1.94\ \text{keV}$. 
In this case, the uncertainty principle does not provide a useful restriction on 
the angle, but low-energy, forward-scattering experiments with neutrons
are certainly possible nonetheless. The authors of Ref.~\cite{Bowman:2014fca}
note that it should be possible to detect a scattering angle of $0.003\ \text{rad}$, and
for definiteness we employ this angle for our low-energy $n-p$ cross section estimate. 
Herewith we summarize our $n-e$ and $n-p$ cross section results. For 
$|\pmb{p}_n| = 0.447 \ \text{GeV}$ we have 
\begin{eqnarray}
\sigma_{n-e}=\left[ |\tilde \delta|^2 \ 5.74 \times 10^7 \right] \ \text{GeV}^{-2} \quad ; \quad 
\sigma_{n-p}=\left[ |\tilde \delta|^2 \ 3.96 \times 10^2 \right] \ \text{GeV}^{-2} \,,
\end{eqnarray}
whereas for $|\pmb{p}_n| = 1.94 \ \text{keV}$ we have 
\begin{eqnarray} 
\sigma_{n-e}= \left[ |\tilde \delta|^2  0.881 \times 10^{21} \right] \ 
\text{GeV}^{-2} \quad ; \quad 
\sigma_{n-p}= \left[ |\tilde \delta|^2 
\ 1.98 \times 10^{13} \right] \ \text{GeV}^{-2} \,.
\end{eqnarray}
The $n-e$ and low-energy $n-p$ results should be regarded as lower bounds. 
Nevertheless, it is apparent that the effects of $n-p$ interactions are relatively 
negligible, and we ignore them in our sensitivity estimates to follow. 
For the cold neutron case, we employ the beam parameters of the ILL 
experiment~\cite{baldo1994new}, so that 
we use $\phi \simeq 1.7 \times 10^{11} \ \text{s}^{-1}$ in our estimate. 
For the higher energy case, we note the study of 
 high-energy (1-120 MeV) neutron 
flux spectra at the Spallation Neutron Source (SNS)
in Fig.\ 5.12 of Ref.~\cite{luciano}, so that 
we employ a flux of $\phi = 5\times 10^8 \ \text{s}^{-1}$ in that case. 
Thus for $|\pmb{p_n}|= 0.447\ \text{GeV}$
the sensitivity to $|\tilde \delta|$ is 
\begin{equation} 
|\tilde \delta| \lesssim 2 \times 10^{-11} 
\sqrt{\frac{N\ \text{events}}{1\ \text{event}}}
\sqrt{\frac{1\ \text{yr}}{\text{t}}}
\sqrt{\frac{5 \times 10^{8}\ \text{s}^{-1}}{\phi}}
\sqrt{\frac{1\ \text{m}}{L}}\sqrt{\frac{5\times 10^{22}\ \text{cm}^{-3}}{\rho}}\ \text{GeV}\,,
\end{equation}
whereas for $|\pmb{p_n}|= 1.94\ \text{keV}$, we have 
\begin{equation}
|\tilde \delta| \lesssim  3 \times 10^{-19} 
\sqrt{\frac{N\ \text{events}}{1\ \text{event}}}
\sqrt{\frac{1\ \text{yr}}{\text{t}}}
\sqrt{\frac{1.7 \times 10^{11}\ \text{s}^{-1}}{\phi}}
\sqrt{\frac{1\ \text{m}}{L}}\sqrt{\frac{5\times 10^{22}\ \text{cm}^{-3}}{\rho}}\ \text{GeV} \,. 
\end{equation} 
Finally, we turn to the case of a solid $^{16}$O target, for which the 
at 24 K is $\rho_o=5.76\times10^{22} \text{cm}^{-3}$~\cite{roder}.  Here, too, 
we focus on the $n-e$ scattering contribution. Since each O atom has eight electrons, 
the cross section should be eight times larger. Thus for 
$|\pmb{p_n}|= 0.447\ \text{GeV}$ we estimate a sensitivity of 
\begin{equation}
|\tilde \delta| \lesssim 7 \times 10^{-12} 
\sqrt{\frac{N\ \text{events}}{1\ \text{event}}}
\sqrt{\frac{1\ \text{yr}}{\text{t}}}
\sqrt{\frac{5 \times 10^{8}\ \text{s}^{-1}}{\phi}}
\sqrt{\frac{1\ \text{m}}{L}}\sqrt{\frac{5.76\times 10^{22}\ \text{cm}^{-3}}{\rho}}\ \text{GeV} \,,
\end{equation}
whereas for $|\pmb{p_n}|= 1.947\ \text{keV}$ we have 
\begin{equation}
|\tilde \delta| \lesssim 1 \times 10^{-19} 
\sqrt{\frac{N\ \text{events}}{1\ \text{event}}}
\sqrt{\frac{1\ \text{yr}}{\text{t}}}
\sqrt{\frac{1.7 \times 10^{11}\ \text{s}^{-1}}{\phi}}
\sqrt{\frac{1\ \text{m}}{L}}\sqrt{\frac{5.76\times 10^{22}\ \text{cm}^{-3}}{\rho}}\ \text{GeV} \,. 
\end{equation} 
It appears that the greatest sensitivity to the $n$-${\bar n}$ oscillation parameter
$\tilde \delta$ can be realized through cold neutron beams scattering from atomic
(or molecular) deuterium or $^{16}$O targets. We wish to emphasize that these particular 
estimates rely on choosing the largest value of a very small scattering angle, 
supposing that the detection of annihilation events would rely on their displacement away 
from the forward direction. 
We note that the measurement of much smaller momentum transfers in neutron 
scattering than we have considered are under development~\cite{vsans}, and 
the realization of this could ultimately lead to significant improvements in sensitivity. 
Improvements 
could also come from the use of brighter neutron beams. 
This could be realizable, e.g., with the planned LD$_2$ cold source for the NG-C guide at
NIST, where we refer to Fig.\ 8 in Ref.~\cite{Cook:2009} for further details. 
The best prospects in this regard, 
however, should be offered by the European Spallation Source (ESS), noting
Refs.~\cite{phillips2016neutron,milstead2015new} for a description of the possibilities.

\section{Summary and Outlook}
\label{outlook}
In this paper we have considered the process of $n$-${\bar n}$ conversion, 
in which a $|\Delta B|=2$ transition, which breaks B-L symmetry, 
is mediated by an external source.  The observation of such a process would reveal 
the existence of physics beyond the SM and indeed that of fundamental Majorana dynamics. 
In contradistinction to $n$-${\bar n}$ oscillation, in which a neutron 
spontaneously converts into an antineutron, the process is not sensitive to 
the presence of fields and matter in the external environment because 
energy-momentum conservation is ensured through the participation of an external current. 
We have developed the connections between $n$-${\bar n}$ conversion and 
$n$-${\bar n}$ oscillation, noting, in particular, that operators that give
rise to spontaneous $n$-${\bar n}$ transitions can also give rise to 
$n$-${\bar n}$ conversion via an external electromagnetic current, because the quarks 
carry electric charge. We have determined precisely how quark-level 
conversion operators can be determined in this case 
and, moreover, how only certain of the operators that generate 
$n$-${\bar n}$ oscillation can also generate $n$-${\bar n}$ conversion. 
Thus searches for  $n$-${\bar n}$ conversion 
are genuinely complementary to  those 
for $n$-${\bar n}$ oscillation. 

We have also studied the inferred limits on the subset of 
low-energy constants associated with $n$-${\bar n}$ oscillation that could arise from 
$n$-${\bar n}$ conversion 
searches. We have found that the connection is sharpest when the momentum transfer
associated with the scattering is smallest, so that the higher-mass-dimension conversion
operator is the least suppressed. We have, moreover,  
evaluated a number of electron-neutron scattering processes and
have found that cold neutron beams scattering from atomic (or molecular) deuterium or
oxygen targets appear to have the greatest sensitivity. 
Generally our anticipated limits are much less severe than those associated with direct 
searches for $n$-${\bar n}$ oscillation, recalling that
the free neutron limit from the ILL experiment can be expressed as 
$\delta \le 5 \times 10^{-32}\,\hbox{GeV}$ at 90\% CL~\cite{baldo1994new}, 
but the set of probed operators is different. 
Also a quantitative  understanding of the manner in which the spontaneous process 
is suppressed by external fields and matter is necessary to assessing those limits. 
The study of B-L violation in scattering does not have such a liability, and 
we note the prospects for the discovery of B-L 
violation via $n$-${\bar n}$ conversion without reference to $n$-${\bar n}$ oscillation 
in a separate paper~\cite{svgxy_search17}. 

Finally we would like to note that the mechanism of $n$-${\bar n}$ conversion 
can lead to broader studies of the spin and flavor 
dependence of B-L violating processes. 
We note the 
prospect of $\Delta^{0}-\bar{\Delta}^{0}$ transitions, as well as that 
of $n-\bar{\Delta}^0$, although the quark-level operators that appear are shared by 
$n$-${\bar n}$ conversion. It is also possible to probe B-L violating operators
that also change strangeness, mediating, e.g., $n-\bar{\Lambda}$ and $n-\bar{\Sigma}^0$
transitions~\cite{basecq1983deltab}, or $\Lambda_0-{\bar \Lambda}_0$ 
transitions~\cite{Kang:2009xt,Addazi:2017iho}.
More generally, we note that the ongoing technical efforts in the 
realization of the next generation of high-intensity electron and neutron beams 
can have an immediate impact on fundamental physics through searches for B-L 
violation via $n$-${\bar n}$ conversion.

\appendix
\section*{Appendices}
\renewcommand{\thesubsection}{\Alph{subsection}}

\subsection{Definitions and conventions}
\label{CPTdef}
In this appendix we collect the definitions and basic results that underlie 
the central arguments of the paper. 
The discrete-symmetry transformations of a four-component fermion field $\psi(x)$ are given by 
\begin{eqnarray}
&&\mathbf{C}\psi(x)\mathbf{C}^{-1} = \eta_c C \gamma^0 \psi^{\ast}(x) \equiv 
\eta_{c}i\gamma^{2}\psi^{\ast}(x) \equiv \eta_c \psi^c (x) \,,\label{Cdef}\\
&&\mathbf{P}\psi(t,\mathbf{x})\mathbf{P}^{-1}=\eta_{p}\gamma^{0}\psi(t,-\mathbf{x}) \,, 
\label{Pdef}\\
&&\mathbf{T}\psi(t,\mathbf{x})\mathbf{T}^{-1}=\eta_{t}\gamma^{1}\gamma^{3}\psi(-t,\mathbf{x})\,,
\label{Tdef}
\end{eqnarray}
where $\eta_{c}$, $\eta_{p}$, and $\eta_{t}$ are unimodular phase factors of the 
charge-conjugation C, parity P, and time-reversal 
T transformations, respectively, and we have chosen the 
Dirac-Pauli representation for the gamma matrices. 
Furthermore the unimodular factors are constrained so that 
$\eta_c\eta_p\eta_t$ and $\eta_p$ are pure imaginary~\cite{gardner2016c}. 

The plane-wave expansion of a Dirac field $\psi(x)$ (noting $\hbar=c=1$) is given by
\begin{equation}
\psi(x) = \int \frac{d^3 \mathbf{p}}{(2\pi)^{3/2} \sqrt{2 E}}
\sum_{s=\pm} \left\{ b(\mathbf{p}, s) u(\mathbf{p}, s) e^{-ip\cdot x}
+ d^\dagger(\mathbf{p}, s) v(\mathbf{p}, s) e^{ip\cdot x}
\right\} \,,
\label{free} 
\end{equation}
with spinors defined as 
\begin{equation}
u(\mathbf{p}, s) = {\cal N}
\left(\begin{array}{c}
\chi^{(s)} \\
\frac{\mathbf{\sigma}\cdot \mathbf{p}}{E_p + M} \chi^{(s)} \\
\end{array}\right) 
\quad ; \quad
v(\mathbf{p}, s) = {\cal N}
\left(\begin{array}{c}
\frac{\mathbf{\sigma}\cdot \mathbf{p}}{E_p + M} \chi^{\prime\, (s)} \\
\chi^{\prime (s)} \\
\end{array}\right) \,,
\end{equation} 
noting $\chi^{\prime\, (s)} = -i \sigma^2 \chi^{(s)}$, 
$\chi^{+} = \left( \stackrel{1}{{}_0} \right)$,  
$\chi^{-} = \left(\stackrel{0}{{}_1} \right)$, and ${\cal N}=\sqrt{E_p + M}$. 
The pertinent fermion anticommutation relations are 
$\{ b(\mathbf{p}, s), b^\dagger(\mathbf{p'}, s') \} =  \{ d(\mathbf{p}, s), d^\dagger(\mathbf{p'}, s') \} = \delta^{(3)}(\mathbf{p} - \mathbf{p'}) \delta^{s\,s'}$; 
all others vanish. We also give the single-particle states a covariant normalization; e.g.,
\begin{equation} 
| n(\mathbf{p}, s) \rangle =\sqrt{2 E_p} (2\pi)^{3/2} b^\dagger (\mathbf{p}, s) | 0 \rangle  \,. 
\end{equation}
With these choices we recover the usual form of the equal-time commutation relations in $\psi(x)$ and $\bar\psi(x)$ and 
that of $\langle n(\mathbf{p}, s) | n(\mathbf{p'}, s') \rangle$~\cite{Peskin:1995ev}. 

We also note the convenient relationships 
\begin{equation}
v^T (\mathbf{p}, s) C = {\bar u}(\mathbf{p}, s) \quad ; \quad 
u^T (\mathbf{p}, s) C = {\bar v}(\mathbf{p}, s)\;, 
\end{equation}
as well as 
\begin{eqnarray}
\bar{u}(\mathbf{p}, s) \gamma^\mu u(\mathbf{p}^\prime, s^\prime)
&=&  \bar{v}(\mathbf{p}^\prime, s^\prime) \gamma^\mu  v(\mathbf{p}, s) \,, \\
\bar{u}(\mathbf{p}, s) \gamma^\mu \gamma_5 u(\mathbf{p}^\prime, s^\prime)
&=& - \bar{v}(\mathbf{p}^\prime, s^\prime) \gamma^\mu \gamma_5 v(\mathbf{p}, s) \,,
\end{eqnarray}
where the latter follow from computing the transpose of the left-hand side in each case. 

\subsection{
The ground-state antiparticle in the M.I.T. bag model}
\begin{figure}
\centering
\includegraphics[scale=0.6]{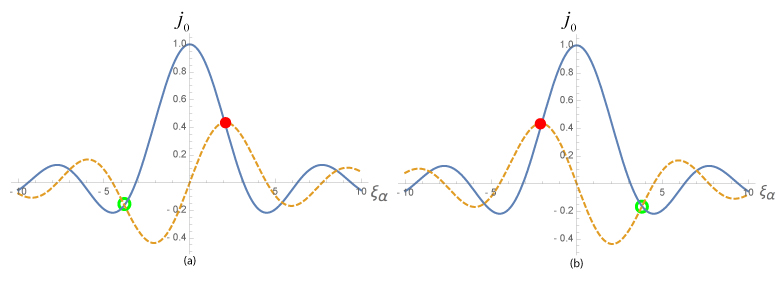}
\caption{Graphical illustration of the solutions to the eigenvalue equation of Eq.~(\ref{eigenxi})
for a ground-state (a) quark,
which has $k=-1$,  and (b) antiquark, which has $k=1$. Note that the solid curve denotes
$j_0$ in each case, whereas the dashed line 
is ${j_1}/{\epsilon_\alpha}$ in (a) and ${-j_1}/{\epsilon_\alpha}$ in (b). The solid
red dot shows the proper solution in each case. 
} 
\label{fig:graph}
\end{figure}

 In the M.I.T. bag model, the 
 quarks and antiquarks obey the free-particle Dirac equation within a 
spherical cavity of radius $R$, subject to boundary conditions at its surface~\cite{Chodos:1974pn,Chodos:1974je}. 
Solutions of opposite parity, i.e., with $k=\mp 1$  exist: 
\begin{eqnarray}
\psi_{\alpha(k=-1)}(\mathbf{r})&=&\frac{N_{\alpha}}{\sqrt{4\pi}}\begin{pmatrix}
ij_0(p_{\alpha}\mathbf{r})\chi^{(s)} \\ -\epsilon_{\alpha} j_1(p_{\alpha}\mathbf{r})\mathbf{\sigma}\cdot \hat{\mathbf{r}} \chi^{(s)}
\end{pmatrix}, \\
\psi_{\alpha(k=1)}(\mathbf{r})&=&\frac{{\tilde N}_{\alpha}}{\sqrt{4\pi}}\begin{pmatrix}
i j_1(p_{\alpha}\mathbf{r})\mathbf{\sigma}\cdot \hat{\mathbf{r}} \chi^{(s)} 
\\ \epsilon_{\alpha}j_0(p_{\alpha}\mathbf{r}) \chi^{(s)}
\end{pmatrix},
\end{eqnarray}
where 
\begin{eqnarray}
N_{\alpha}&=&\Big(\frac{\xi_{\alpha}^2j^{-2}_0(\xi_{\alpha})}{R^3[2E_{\alpha}R(E_{\alpha}R-1)+ m R]}\Big)^{\frac{1}{2}},\\
{\tilde N}_{\alpha}&=&\Big(\frac{\xi_{\alpha}^2j^{-2}_1(\xi_{\alpha})}{R^3[2E_{\alpha}R(E_{\alpha}R+1)+ m R]}\Big)^{\frac{1}{2}} \,,
\end{eqnarray}
and $j_n$ is a spherical Bessel function of the first kind of order $n$, with $m$ and $s$ 
denoting the quark mass and spin, respectively. Also 
\begin{equation}
p_{\alpha}=\frac{\xi_{\alpha}}{R} \,, \quad  E^2_{\alpha}= p^2_{\alpha} + m^2 \, , \quad 
\epsilon_{\alpha}=\sqrt{\frac{E_{\alpha}-m}{E_{\alpha}+m}}\,,
\end{equation} 
where an eigenvalue equation
determines the quantity $\xi_{\alpha}$ and eventually the energy $E_{\alpha}$, namely, 
\begin{eqnarray}
j_1(\xi_{\alpha})=\pm\epsilon_{\alpha}j_0(\xi_{\alpha}) \,,
\label{eigenxi}
\end{eqnarray}
or, 
 equivalently, 
\begin{eqnarray}
\tan \xi_{\alpha}=\frac{k\xi_{\alpha}}{k-km R+E_{\alpha}R}\,
\end{eqnarray}
for $k=-1 \,(+)$ or $k=1 \,(-)$. The prescription for a ground-state 
antiquark has not been clearly stated~\cite{Shrock:1978dm}, and mistakes have appeared 
in the literature~\cite{pasupathy1982neutron}. Here we clarify the proper choice 
through Fig.~\ref{fig:graph}, which illustrates the solutions to 
the eigenvalue equation for $k=\mp 1$. The ground-state solution for a quark is given by the solid
red dot in Fig.~\ref{fig:graph}a. 
In order for the magnitude of the energy of the ground-state antiquark to be same as that of the antiquark 
we must also pick the solution marked by the solid red dot in Fig.~\ref{fig:graph}b. 
Consequently the quark and antiquark solutions are related by 
\begin{eqnarray}
\bar{\xi}_{\alpha}&=&-\xi_{\alpha},\\
\bar{E}_{\alpha}&=&-E_{\alpha},
\end{eqnarray} 
where  ${\bar \xi}_\alpha$ and ${\bar E}_\alpha$ denote those of the 
ground-state antiparticle. The quark state is $u_{\alpha , 0}^s (\mathbf{r}) = \psi_{\alpha (k=-1)}(\mathbf{r})$. 
In contrast, the true solution for an antiquark $v_{\alpha, 0}^s (\mathbf{r})$ should be

\begin{eqnarray}
v_{\alpha, 0} (\mathbf{r})&=&\frac{\bar{N}_{\alpha}}{\sqrt{4\pi}}\begin{pmatrix}
-i j_1(p_{\alpha}\mathbf{r})\mathbf{\sigma}\cdot \hat{\mathbf{r}} \chi^{\prime (s)} \\ \bar{\epsilon}_{\alpha}j_0(p_{\alpha}\mathbf{r}) \chi^{\prime (s)}
\end{pmatrix} \,,
\end{eqnarray}
where
\begin{eqnarray}
\bar{N}_{\alpha}&=&\Big(\frac{\xi_{\alpha}^2j^{-2}_1(\xi_{\alpha})}{R^3[-2E_{\alpha}R(-E_{\alpha}R+1)+m R]}\Big)^{\frac{1}{2}}\nonumber \\
&=&\Big(\frac{\xi_{\alpha}^2j^{-2}_0(\xi_{\alpha})}{R^3[2E_{\alpha}R(E_{\alpha}R-1)+ m R]}\Big)^{\frac{1}{2}}\bar{\epsilon}_{\alpha}^{-1},
\end{eqnarray}
with
\begin{eqnarray}
 \bar{\epsilon}_{\alpha}=\sqrt{\frac{-E_{\alpha}-m}{-E_{\alpha}+m}}=\epsilon_\alpha^{-1}.
\end{eqnarray}
Put everything together we have 
\begin{eqnarray}
v_{\alpha, 0}^s (\mathbf{r}) &=&\frac{N_{\alpha }}{\sqrt{4\pi}}\begin{pmatrix}
i\epsilon_\alpha j_1(p_{\alpha}\mathbf{r})\mathbf{\sigma}\cdot \hat{\mathbf{r}} \chi^{\prime\,(s)} \\ -j_0(p_{\alpha}\mathbf{r}) \chi^{\prime\,(s)}
\end{pmatrix}.
\end{eqnarray}

\begin{acknowledgments}
We acknowledge partial support from the U.S. Department
of Energy Office of Nuclear Physics under 
contract DE-FG02-96ER40989, and we thank our colleagues, particularly 
Jonathan Feng, 
at the University of California, Irvine (UCI) for generous hospitality. 
X.Y. would also like to thank the Graduate School of the 
University of Kentucky and the Huffaker Fund for providing travel support to UCI. 
S.G. is also grateful to the Mainz for Institute Theoretical Physics for partial 
support and gracious hospitality during the final phases of this project. 
We also thank Robert Jaffe for key correspondence 
regarding antiquark states in the M.I.T. bag model 
and Christopher Crawford, Geoffrey Greene, Wick Haxton, 
Shannon Hoogerheide, Wolfgang Korsch, Kent Leung, Bradley Plaster, 
and Michael Snow for helpful comments and/or references. 
\end{acknowledgments}

\bibliography{nnbar_conversion}
\end{document}